\documentclass[preprint]{aastex62}
\usepackage[USenglish]{babel}
\usepackage[utf8]{inputenc}
\usepackage[T1]{fontenc}
\usepackage{newtxtext,newtxmath}
\usepackage{xspace}
\usepackage{wrapfig}

\graphicspath{{./figs/}}

\begin{document}
\title{Astrophysical Tests of Dark Matter with Maunakea Spectroscopic Explorer}

\author{Ting~S.~Li}
\affiliation{Fermi National Accelerator Laboratory}
\affiliation{Kavli Institute of Cosmological Physics, University of Chicago}
\author{Manoj~Kaplinghat}
\affiliation{Department of Physics and Astronomy, University of California, Irvine, CA 92697, USA}
\author{Keith~Bechtol}
\affiliation{Physics Department, University of Wisconsin-Madison}
\author{Adam~S.~Bolton}
\affiliation{National Optical Astronomy Observatory, 950 N Cherry Ave, Tucson, AZ 85719 USA}
\author{Jo~Bovy}
\affiliation{Department of Astronomy \& Astrophysics, University of Toronto, Canada}
\author{Timothy~Carleton}
\affiliation{Department of Physics and Astronomy, University of Missouri, Columbia, MO  65211}
\author{Chihway~Chang}
\affiliation{Department of Astronomy \& Astrophysics, University of Chicago}
\affiliation{Kavli Institute of Cosmological Physics, University of Chicago}
\author{Alex~Drlica-Wagner}
\affiliation{Fermi National Accelerator Laboratory}
\affiliation{Kavli Institute of Cosmological Physics, University of Chicago}
\affiliation{Department of Astronomy \& Astrophysics, University of Chicago}
\author{Denis~Erkal}
\affiliation{Department of Physics, University of Surrey, UK}
\author{Marla~Geha}
\affiliation{Department of Astronomy, Yale University, New Haven, CT 06520, USA}
\author{Johnny~P.~Greco}
\affiliation{Center for Cosmology and AstroParticle Physics, The Ohio State University}
\author{Carl~J.~Grillmair}
\affiliation{IPAC, California Institute of Technology, Pasadena, CA 91125}
\author{Stacy~Y.~Kim}
\affiliation{Department of Astronomy, The Ohio State University}
\author{Chervin~F.~P.~Laporte}
\affiliation{CITA National Fellow, Department of Physics and Astornomy, University of Victoria, 3800 Finnerty Road, Victoria, BC, V8P 5C2, Canada}
\author{Geraint~F.~Lewis}
\affiliation{Sydney Institute for Astronomy, School of Physics, A28, The University of Sydney, NSW 2006, Australia}
\author{Martin~Makler}
\affiliation{Brazilian Center for Physics Research, Rio de Janeiro, RJ 22290-180, Brazil}
\author{Yao-Yuan~Mao}
\affiliation{Department of Physics and Astronomy and Pittsburgh Particle Physics, Astrophysics and Cosmology Center (PITT PACC), University of Pittsburgh}
\author{Jennifer~L.~Marshall}
\affiliation{George P. and Cynthia Woods Mitchell Institute for Fundamental Physics and Astronomy, and Department of Physics and Astronomy, Texas A\&M University, College Station, TX 77843,  USA}
\author{Alan~W.~McConnachie}
\affiliation{NRC Herzberg Astronomy and Astrophysics, 5071 West Saanich Road, Victoria, BC V9E 2E7, Canada}
\author{Lina~Necib}
\affiliation{Walter Burke Institute for Theoretical Physics, California Institute of Technology, Pasadena, CA 91125, USA}
\author{A.~M.~Nierenberg}
\affiliation{Jet Propulsion Laboratory, 4800 Oak Grove Dr., Pasadena CA 91109 }
\author{Brian~Nord}
\affiliation{Fermi National Accelerator Laboratory}
\affiliation{Kavli Institute of Cosmological Physics, University of Chicago}
\affiliation{Department of Astronomy \& Astrophysics, University of Chicago}
\author{Andrew~B.~Pace}
\affiliation{George P. and Cynthia Woods Mitchell Institute for Fundamental Physics and Astronomy, and Department of Physics and Astronomy, Texas A\&M University, College Station, TX 77843,  USA}
\author{Marcel~S.~Pawlowski}
\affiliation{Department of Physics and Astronomy, University of California, Irvine, CA 92697, USA}
\affiliation{Leibniz-Institut f\"ur Astrophysik Potsdam (AIP), An der Sternwarte 16, D-14482 Potsdam, Germany}
\author{Annika~H.~G.~Peter}
\affiliation{Department of Physics, The Ohio State University}
\affiliation{Center for Cosmology and AstroParticle Physics, The Ohio State University}
\affiliation{Department of Astronomy, The Ohio State University}
\author{Robyn~E.~Sanderson}
\affiliation{Department of Physics \& Astronomy, University of Pennsylvania, 209 S 33rd St, Philadelphia, PA 19104 USA}
\affiliation{Center for Computational Astrophysics, Flatiron Institute, 162 5th Ave, New York, NY 10010}
\author{Guillaume~F.~Thomas}
\affiliation{NRC Herzberg Astronomy and Astrophysics, 5071 West Saanich Road, Victoria, BC V9E 2E7, Canada}
\author{Erik~Tollerud}
\affiliation{Space Telescope Science Institute, Baltimore, MD 21211, USA}
\author{Simona~Vegetti}
\affiliation{Max-Planck Institut f\"ur Astrophysik, Karl Schwarzschildstrasse 1, D-85748 Garching, Germany}
\author{Matthew~G.~Walker}
\affiliation{McWilliams Center for Cosmology, Carnegie Mellon University}
\affiliation{Department of Physics, Carnegie Mellon University}


\correspondingauthor{{\bf Ting~S.~Li}, {\bf Manoj~Kaplinghat}} \email{tingli@fnal.gov, mkapling@uci.edu}

\section*{Abstract}
We discuss how astrophysical observations with the Maunakea Spectroscopic Explorer (MSE), a high-multiplexity (about 4300 fibers), wide field-of-view (1.5 square degree), large telescope aperture (11.25 m) facility, can probe the particle nature of dark matter. MSE will conduct a suite of surveys that will provide critical input for determinations of the mass function, phase-space distribution, and internal density profiles of dark matter halos across all mass scales. N-body and hydrodynamical simulations of cold, warm, fuzzy and self-interacting dark matter suggest that non-trivial dynamics in the dark sector could have left an imprint on structure formation. Analysed within these frameworks, the extensive and unprecedented datasets produced by MSE will be used to search for deviations away from cold and collisionless dark matter model. MSE will provide an improved estimate of the local density of dark matter, critical for direct detection experiments, and will improve estimates of the J-factor for indirect searches through self-annihilation or decay into Standard Model particles. MSE will determine the impact of low mass substructures on the dynamics of Milky Way stellar streams in velocity space, and will allow for estimates of the density profiles of the dark matter halos of Milky Way dwarf galaxies using more than an order of magnitude more tracers. In the low redshift Universe, MSE will provide critical redshifts to pin down the luminosity functions of vast numbers of satellite systems, and MSE will be an essential component of future strong lensing measurements to constrain the halo mass function. Across nearly all mass scales, the improvements offered by MSE, in comparison to other facilities, are such that the relevant analyses are limited by systematics rather than statistics.

\newpage

\tableofcontents

\newpage

\section{Motivation}\label{sec:motivation}

Dark matter has been detected through its gravitational influence on galaxies and clusters of galaxies, the large-scale distribution of galaxies and the cosmic microwave background. A cosmological model with particle dark matter convincingly explains a vast array of observations stretching from kiloparsec scales to the horizon and from the present time to the time of last scattering \citep{Davis1985}.  

Dark matter density equivalent to about $0.3 \rm GeV/cc$ has been inferred in the solar neighborhood from the motions of disk stars \citep{Kuijken91,HF04,garbari12,2012ApJ...756...89B}. We know from the orbit of the Milky Way and Andromeda that the two combined have a mass of about  $2\times 10^{12} \mathrm{M}_\odot$, far in excess of all the stars and gas \citep[][]{Kahn1959, Penarrubia2014}. The satellite galaxies orbiting the Milky Way provide strong evidence for dark matter, with inferences of 10 to 1000 times more mass in dark matter than stars \citep{Mateo1998, sg07, 2008Natur.454.1096S}. Away from our Local Group, every galaxy for which we have dynamical information far enough out of the disk of stars has shown evidence for dark matter. Amazingly, all these measurements are consistent within a factor of two with the predictions of hierarchical structure formation models with dark matter.

In larger systems like groups and clusters of galaxies, we see concrete evidence for dark matter through different methods, with densities that scale in the way expected from dark matter models \citep{Navarro1997}. The large-scale distribution of galaxies stretching over hundreds of Mpc is beautifully explained in the context of a model which includes dark matter \citep[e.g.][]{Davis1985,Springel2005b}. On even larger scales and from the time when the Universe was about four hundred thousand years old, we have clear evidence for non-baryonic (dark) matter in the cosmic microwave background (CMB) anisotropies \citep{komatsu09,Planck18}.

While the model space of dark matter is large \citep{2010ARA&A..48..495F}, the dominant idea in the particle physics community has been that the dark matter particle is the lightest supersymmetric (SUSY) particle (a ``neutralino") or the axion. Both candidates have the virtue that they arose in models designed to solve deep problems in particle physics. However, the neutralino or the axion does not have to be the dominant component of dark matter. 

Despite enormous progress in mapping the distribution of dark matter in galaxies and a concerted effort to look for certain kinds of dark matter particles in underground laboratories, at colliders and in space, there is no concrete evidence for the identity of the dark matter particle. The lack of detection has ruled out large parts of parameter space \citep[e.g.][]{Cohen:2013ama,Aprile:2018dbl,2018EPJC...78..203A} and pushed theorists to explore more general models of dark matter \citep[e.g.][]{apsarticle}. Many of the models can be broadly classified as dark sector models, i.e., models where the dark matter lives in a secluded sector that is very weakly coupled to the Standard Model of particle physics.  
Within this theory landscape, there are many ideas being currently explored \citep{2017arXiv170704591B}.

In the dark sector, there is no 
compelling 
reason to expect the dark matter particle mass to be $\mathcal{O}(100\, \rm GeV/c^2)$
(i.e., weak scale mass). 
Just like Standard Model particles can be light and have appreciable interactions via forces other than gravity, the dark matter in the hidden sector can also be light (sterile neutrino dark matter \citealt{dodelson1994, shi1999}; fuzzy dark matter, \citealt{Hu2000, hui2017}) and have large interactions, including interactions with itself (for example, like Hydrogen atoms, \citealt{kaplan2010}). There may be ways for the dark matter particles to also cool via inelastic interactions (double disk dark matter, \citealt{fan2013}; atomic dark matter, \citealt{cyr-racine2013, foot2014}). Many of these exciting possibilities can only be tested by astrophysical probes (\citealt{boddy2016, Vogelsberger2018}).

Stepping away from the theoretical landscape, there are other reasons to look towards the dark sector. A long standing puzzle in the galaxy formation community has been the presence of spiral and dwarf galaxies with low dark matter densities in the center. This issue is often referred to as the cusp-core problem. 
Halos formed in N-body simulations with cold collisionless dark matter (CDM) have central ``cusps'' such that the density increases with decreasing radius ($r$) as $1/r$, whereas rotation curves and stellar kinematics of many galaxies show evidence for ``cores'' of uniform density \citep{Moore1994, Flores1994, deBlok2010, Walker:2011}.
However, there are also galaxies with similar total baryon content that have densities similar to the CDM predictions. Excitingly, this diversity seems to be consistent with predictions of models where dark matter has large self interactions~\citep{2017PhRvL.119k1102K}. 

At the faint end, it becomes harder to decipher cores and cusps but a similar problem, referred to as the ``too big to fail" problem, exists~\citep{Boylan-Kolchin2012,papastergis2015}. Solutions to these problems in terms of dark matter physics are more varied and can include dark matter self interactions, warm dark matter or fuzzy dark matter.

It is important to note that just the  presence of low-density cores in galaxies is not evidence for deviations from the CDM model. It has become clear in recent years that large cores can be created in the context of CDM models with better feedback prescriptions. \citep[e.g.][]{Brooks2013, Wetzel2016,2016MNRAS.456.3542T}. This progress underscores the point that we cannot talk about dark matter halo properties in isolation from star formation considerations. There is great diversity of dark matter cores across a wide range of galaxies  \citep{mcgaugh2005,denaray2010,oman2015}, and the cores are correlated with the stellar distribution \citep{mcgaugh2005}. The richness of the cusp-core issue, coupled with the rapid progress in hydrodynamical simulations, indicates that it should be possible to disentangle feedback physics from dark matter physics. 

The case for using astrophysical observables to constrain or measure particle physics models of dark matter is strong. The recent progress in N-body and hydrodynamical simulations of cold, warm, fuzzy and self-interacting dark matter have helped bolster this case, while a wealth of new observations from dwarf galaxies to galaxy cluster scales has opened up the exciting possibility that non-trivial dynamics in the hidden sector could have left an imprint on structure formation. MSE has critical roles to play in this unfolding story and we highlight these below.

\section{How can astrophysics probe the particle nature of dark matter?}\label{sec:justification}

\subsection{Dark matter physics}

One of the fundamental predictions of CDM model is that structure formation is hierarchical \citep{white_rees_1978}, with the smallest structures collapsing first and then merging into larger structures \citep{Davis1985}. 
Indeed, calculations of the matter power spectrum associated with popular ``weakly interacting massive particle” (WIMP; e.g., super-symmetric neutralinos) candidates for the dark matter particle imply that the minimum mass of self-bound structures could be as small as an earth mass~\citep{Hofmann:2001bi, Green2004, Diemand2005,2005PhRvD..71j3520L,2006PhRvD..74f3509B}.
Halos formed in a hierarchical structure formation scenario are predicted to have subhalos, with subhalos hosting their own sub-subhalos down to the scale of the ``minimum" mass set by the particle physics.  To date, the smallest dark matter halos that have been inferred from observations have mass $\sim 10^5M_\odot$ within their luminous regions (often extrapolated to virial masses $\mathrm{M_{vir}} \sim 10^8M_\odot$). These observations thus allow for the possibility that additional physics in the dark sector may inhibit structure formation on smaller scales \citep{Spergel2000, Bode2001, Dalcanton2001,2005PhRvD..72f3510K, Gilmore:2007fy}.

There are two generic ways in which the dark matter particle properties can change the abundance of dark matter halos or their internal density profile. One is an early Universe effect (typically set in place well before matter-radiation equality) and the other a late Universe effect.

In the early Universe, there are two quantities that turn out to be relevant for structure formation. One is the mean velocity of dark matter particles in the early Universe, which sets the free-streaming scale, below which fluctuations are erased. If the dark matter particles have significant interactions with relativistic particles (either dark radiation or interactions with Standard Model leptons or photons), this will lead to an additional suppression of fluctuations through the same mechanism that leads to acoustic oscillations and damping of the cosmic microwave background. The larger of these two effects will determine the suppression scale, $\lambda$ . 

The suppression scale, processed through non-linear structure formation, results in lowered abundance for dark matter halos with mass comparable to or below $(4\pi \lambda^3/3) \rho_{\rm matter}$, where $\rho_{\rm matter}$ is the cosmic abundance of matter today. Halos are not just fewer in number, they also grow later because of the suppression in the power spectrum due to these early Universe effects. The slower growth results in a less concentrated halo.

\begin{figure*}
\centering
  \includegraphics[angle=0,  clip, width=0.6\textwidth]{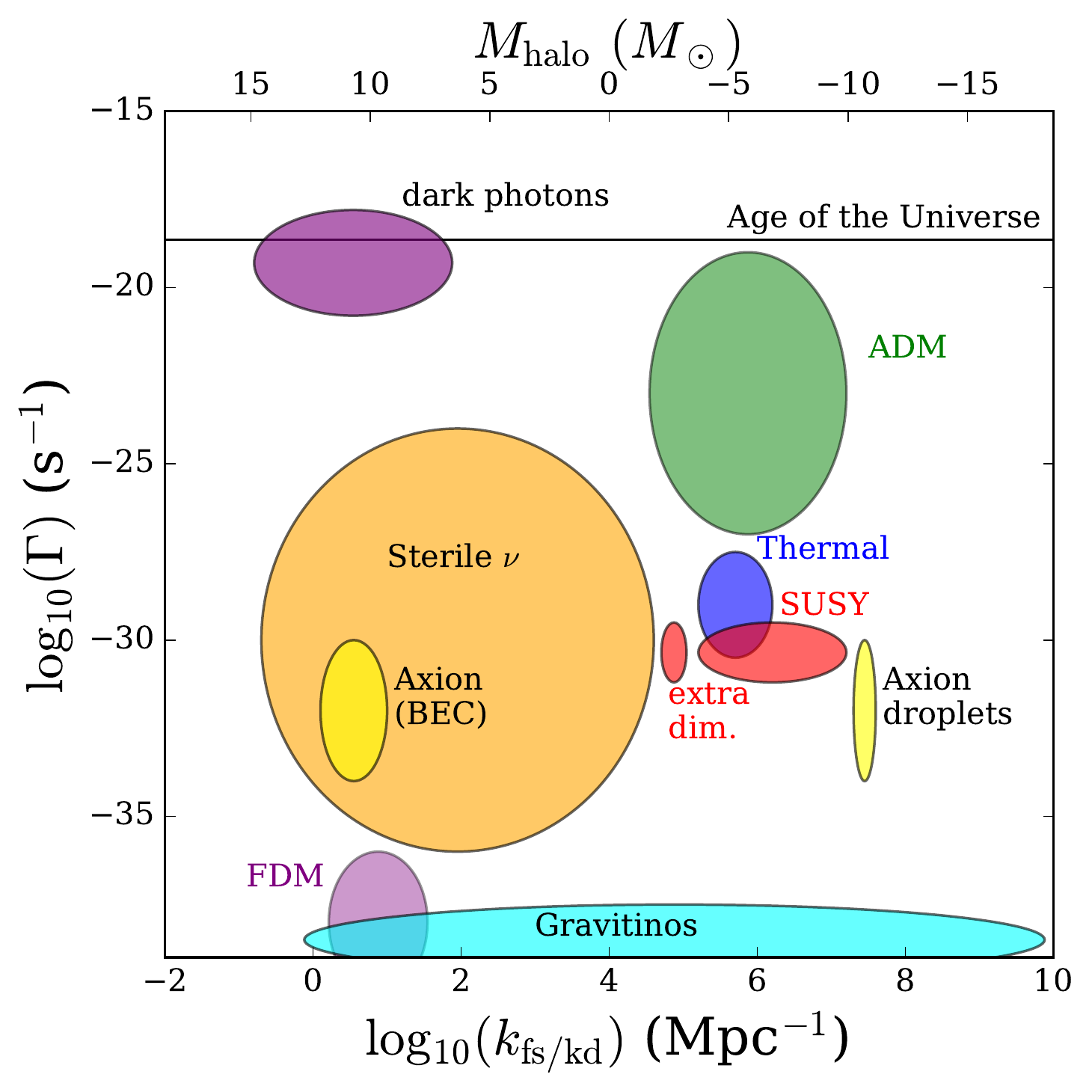}
  \caption{Various dark matter candidates in a parameter space that directly influences structure formation. The horizontal axis is the characteristic free-streaming wavenumber of the model and the corresponding halo mass is shown on the top. These effects are set in the early Universe, typically well before the epoch of matter-radiation equality. The vertical axis is the characteristic interaction or decay rate, which directly impacts the structure of the halos at late times. Figure from \citet{buckleypeter2018}.}
\label{interaction}
\end{figure*}

In addition to the physics described above, one can have dark matter properties such as self-interactions and decay that only impact the halos after their formation (late Universe). Both the self interaction of dark matter particles (elastic or inelastic) and dark matter decay impact a wide range of galaxy masses.

It is not necessary to work out the structure formation of each particle physics model separately.
For astrophysical purposes, dark matter particle candidates and associated cosmologies can generally be classified according to whether the dark matter particle's properties (e.g., mass and corresponding free-streaming scale, non-gravitational interactions, etc.) play a role in galaxy formation~\citep{2016MNRAS.460.1399V}. An example of such a classification is shown in Figure \ref{interaction}. 

For the purposes of delineating the physics, we will use four generic classes of dark matter models, namely CDM, ``warm dark matter" (WDM), ``self-interacting dark matter" (SIDM) and ``fuzzy dark matter" (FDM).

{\bf WDM} refers generically to candidates whose abundance is set in the early Universe when they are relativistic. These particles have a free-streaming scale that is essentially set by their mass ($\gtrsim \mathrm{keV}$). Examples include low-mass gravitinos and sterile neutrinos, which are examples of dark matter candidates arising from minimal models of particle physics. The free-streaming scale introduces a cut-off in the linear power spectrum. The lack of power on small scales implies structure grows later and the minimum halo mass can be as large as that allowed by the data (around $10^8 {\rm M}_\odot$). There is no concrete guidance from theory for why the WDM mass would be in the regime that impacts structure formation; the minimum halo mass could be much smaller and the model would be indistinguishable from CDM in terms of its gravitational imprints. 

{\bf SIDM} refers to dark matter candidates that have appreciable non-gravitational self interactions~\citep{Spergel2000,Vogelsberger2012,Rocha2013}. As an example, imagine a hidden sector with stable neutrons. In this case, the dark matter particles (hidden neutrons) have velocity-dependent self-interactions set by the mass of the hidden sector pion. If the hidden pion and neutron masses are similar to that in the Standard Model, then the self-interaction cross section over mass $\sigma/m$ would of order $10 \, \rm cm^2/g$. Cross sections over mass larger than about $0.1 \, \rm cm^2/g$ are detectable~\citep{2016PhRvL.116d1302K} because they change the density profile of halos and subhalos. The reduced central densities of subhalos can also impact their survival in the tidal field of the parent halos. Other examples of SIDM include hidden H-atoms. In this case, we have the additional process in the early Universe of dark matter particles scattering off of {\em massless} hidden photons \citep{feng2009} and this will lead to dark acoustic oscillations and a cut-off in the linear power spectrum \citep{cyr-racine2013}, mimicking WDM behavior but with the additional late-time phenomenology due to self scattering \citep{buckley2014,boddy2016}. 

{\bf FDM} is another late time effect that has recently been explored through simulations \citep{Hu2000, Hui2016}. In this scenario, dark matter particles have masses below about $10^{-22} \rm eV$ and the effective de Broglie wavelength in galaxies is kpc-sized. Thus the wave nature of dark matter becomes important. The pressure due to the wave nature of dark matter leads to the creation of dense solitonic cores (denser than CDM). The wave nature of dark matter in fuzzy dark matter models also implies a lack of structure on scales below the Jeans length, which is set by the mass $m$ of the (fuzzy) dark matter particle, and the corresponding Jeans mass is roughly $4\times 10^7 M_\odot (10^{-22} \rm eV/m)^{3/2}$ (\citealt{hui2017}). 

\subsection{Observables}

Broadly speaking, there are three aspects of dark matter halos that we would like to infer from observations to constrain or measure dark matter particle properties: the mass function of subhalos, the phase-space distribution of subhalos, and the internal density profiles of field halos and subhalos. A particle physics model can be mapped on to these observables that are amenable to constraints from structure formation.

The {\bf mass function of subhalos} is set, after non-linear processing, by a variety of processes. The linear power spectrum has a direct impact, as discussed before, through the suppression of structure below a threshold halo mass. Processes such as dark matter decay and mass-loss due to self interactions, in conjunction with tidal interactions with the disk and halo of the Milky Way will impact the number of subhalos that survive \citep{D'Onghia10,Errani17}. The survival probability will be a function of the orbital properties such as the pericenter distance and the number of pericenter passages. Thus, the {\bf phase space distribution} of the subhalos (which includes the radial distribution) will be impacted by dark matter physics.

The internal {\bf density profile} of field halos and subhalos has been the other main avenue for constraining dark matter physics. Viable models of WDM do not create cores (the profile retains the $1/r$ cusp) on observable scales \citep{denaray2010,2011JCAP...03..024V,2012MNRAS.424.1105M}. However, the concentration of the halos (and hence the inner density) is lower due to the delayed structure formation \citep{Lovell14}. In SIDM, the density profile is shallower in the center due to heat transfer from the outer to inner parts of the halo~\citep{2001ApJ...547..574D}, but this statement assumes that the core has not started contracting (which is the generic late-time behavior). In FDM, the outer profile is expected to be similar to the CDM profile but in the inner parts a dense solitonic core forms~\citep{2014NatPh..10..496S}, a feature that is disfavored by rotation curve data~\citep{2018PhRvD..98h3027B,2019MNRAS.483..289R}. 

\begin{table*}
\centering
\begin{tabular}{p{7.8cm}|p{1.8cm}|p{1.8cm}|p{1.8cm}|p{1.8cm}}
\hline
    Halo property & WDM & SIDM (massive mediators) & SIDM (massless mediators)& FDM \\
    \hline
    Slope of halo density profile & N & Y & Y & Y\\
    Central density of subhalos and dwarf halos & Y & Y & Y & Y \\
    Central density of more massive halos & N & Y & Y & N \\
    Subhalo mass function & Y & N & Y & Y \\
    \hline
\end{tabular}
\caption{A partial list of how different models of dark matter can impact the halo and subhalo properties in comparison to the CDM predictions. For SIDM, two different examples are included; one in which the dark matter particles interact with a massive  $\mathcal{O}(\rm MeV)$ mediator (e.g., hidden stable neutrons) and one in which the dark matter particles interact with a massless mediator (e.g., atomic dark matter). }\label{dmhalos}
\end{table*}

A few examples will help further elucidate the influence of particle physics on these observable quantities. Table~\ref{dmhalos} provides a concise summary of some of the major effects.

\begin{itemize}
\item 
FDM cuts off halo formation below a mass scale determined by the FDM mass. This could be tested with substructure detections using strong lensing or gaps in stellar streams; 
\item 
SIDM 
transports kinetic energy in halos and subhalos and changes the density profile from the CDM predictions. In halos with small $M_\star/M_{\rm halo}$ (like the Local Group dwarf spheroidal galaxies) this leads to constant density cores (for moderate cross sections), which could be tested with resolved stellar velocities and rotation curves;
\item At the other extreme end, dark matter density profile of the halos of clusters of galaxies provide a sensitive probe of the self interaction cross section at velocities of order 1000\,km\,s$^{-1}$ or larger;
\item SIDM interactions (if large enough) could also evaporate subhalos and change the number of subhalos. The survival of subhalos and their radial distribution in the halo is sensitive to self-interaction strength;
\item Both FDM and SIDM (when it forms constant density cores) change the strength of dynamical friction and this could be a testable prediction. For example, in SIDM models, it is expected that the BCG will slosh about the center of the halo on the scale of the constant density core size. 
\end{itemize}

\subsection{The impact of baryons}

The presence of baryons can change the above description in dramatic ways. Sufficiently vigorous and bursty star formation, strong winds from massive stars and supernova-driven outflows that remove gas rapidly compared with local dynamical timescales can significantly alter the structure of dark matter halos \citep{Navarro:1996b, Read2005, Governato2010, Pontzen2012, Pontzen2014,2016MNRAS.456.3542T,2017MNRAS.471.3547F}. As a result of these influences, the orbits of dark matter particles expand non-adiabatically, potentially transforming central cusps into cores and lowering masses within the half-light radii of dwarf galaxies. If the feedback processes are also important in satellite galaxies (for example, the Fornax dSph), this would lower the binding energies of subhalos, leaving them more vulnerable to tidal disruption analogous to effect of self interactions. 

The impact of feedback on WDM and SIDM has been studied for dwarf galaxies~\citep{2018arXiv181111791F}. For SIDM models in which the cross section is large -- $\sigma/m$ of few $\rm cm^2/g$ -- the drive to thermalization renders the final dark matter density profile insensitive to the star formation history~\citep{2017MNRAS.472.2945R}, but it does depend sensitively on the final distribution of stars and gas~\citep{2014PhRvL.113b1302K}. 

The presence of a disk in the parent halo can have a marked effect on the survival of the subhalos that venture close to the centers of galaxies \citep[e.g., ][]{Brooks2013, Wetzel2016}. At present, however, it is not clear if the disk leads to a divergence in the predictions of these models for the radial distribution of the subhalos or makes the models similar. The answer may depend on the observable of interest, for example, ultrafaint dwarf galaxies or stellar streams may be affected quite differently. 

The promise of precision data across a wide range of scales, from ultrafaint dwarf galaxies deep within the galactic potential to low-surface brightness galaxies in the field to galaxies in the cores of clusters, holds out hope that we can disentangle signatures of the particle nature of dark matter from signatures of feedback from star formation. For example, when feedback is efficient at producing cores in isolated field dwarfs, does it simultaneously also produce high-surface brightness galaxies? When self interactions are efficient at creating cores in isolated low-surface brightness galaxies, do they also match the properties of the dwarf spheroidals of the Milky Way and Andromeda  \citep{collins14, tollerud14}? 

In the remainder of this chapter, we explore the possible ways in which MSE can help to reveal the nature of dark matter. We have grouped the science cases into four sections based on the distance of the stars being targeted: stars and streams in the Milky Way (Section~\ref{sec:streams}); dwarf galaxies in the Milky Way and beyond  (Section~\ref{sec:dwarfs}); galaxies in the low redshift ($z<0.05$) Universe  (Section~\ref{sec:lowz}); galaxies beyond the low redshift Universe  (Section~\ref{sec:highz}).

\section{Stars and stellar streams in the Milky Way}\label{sec:streams}

The Milky Way is a great laboratory for studying the distribution of dark matter on small scales. The ability to  measure 3D positions and 3D velocities for individual stars in the Milky Way implies that we can gravitationally map the dark matter distribution in great detail.
The Milky Way's mass distribution is already well constrained in its inner part \citep[e.g.,][]{2015ApJS..216...29B,McMillan:2017} where much of the mass is baryonic, but constraints on the total mass, radial profile, and shape of the smooth dark-matter distribution remain weak even though these are key observables when comparing against the predictions from different dark matter models. Understanding the smooth dark matter density and velocity distribution better is also necessary for the interpretation of laboratory direct-detection experiments and for indirect-detection probes. Finally, tidal stellar streams in the Milky Way are one of a small number of methods known today for measuring the clustering of dark matter on very small scales ($M \lesssim 10^8\,M_\odot$) by looking for the impact of small dark-matter subhalos on the structure of cold stellar streams. 
Detecting dark subhaloes that do not have any detectable gas or stars would be a stunning confirmation of the dark matter paradigm. 
Detection or constraints on the presence of these subhalos would also provide information about the particle properties of dark matter.

We now describe how, in conjunction with data from Gaia, LSST, and WFIRST, MSE has the ability to transform our knowledge of dark matter in the Milky Way. The essential contribution of MSE to this science is the measurement of line-of-sight velocities to a precision of $1$ to $5\,\mathrm{km\,s}^{-1}$ for extremely large numbers of stellar tracers.

\subsection{Mapping the Milky Way's gravitational potential with stars, dwarf galaxies, and stellar streams}

 \begin{figure*}
\centering
  \includegraphics[angle=0,  clip, width=18cm]{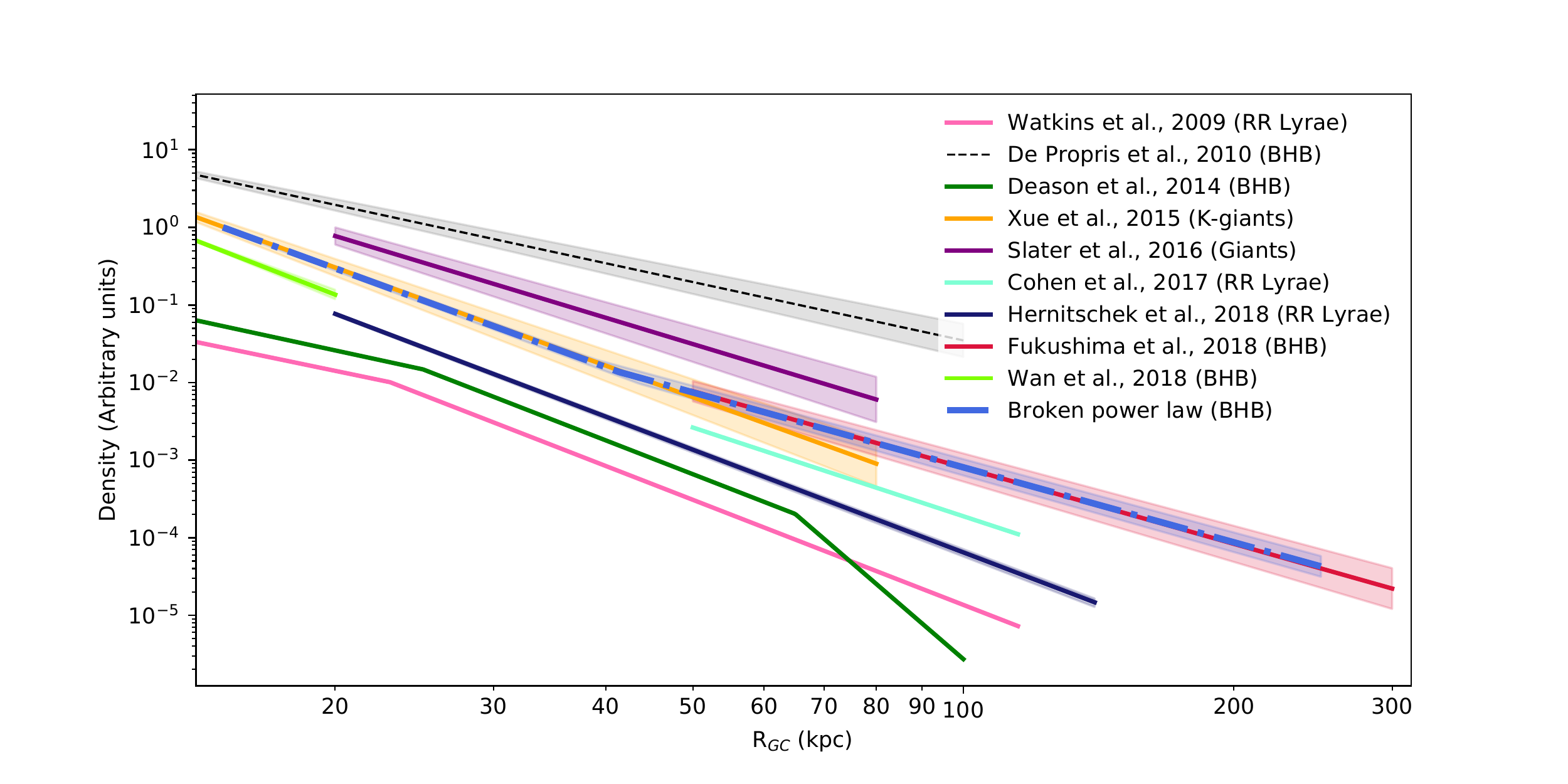}
   \caption{Profile of the outer stellar halo ($R>15$kpc) using different tracers. The typical power slope of the halo is in the range $3 - 4$. However, even for a similar population, such as the Blue Horizontal Branch (BHB) stars, the profile beyond 80 kpc is still very uncertain. Figure from \citet{Thomas2018}.  }
\label{profile_compare}
\end{figure*}

Our understanding of the mass, shape, and mass profile of the Milky Way underlies every effort to test theories of dark matter using the galaxy we know best. The {\it mass} of the Milky Way's halo determines which simulated galaxies we should compare to when assessing consistency of our observations with predictions sensitive to the dark matter model, such as the number and structure of satellite galaxies. The concentration and {\it radial profile} of the Galaxy's dark matter set constraints on its accretion history and formation time \citep{2002ApJ...568...52W}, which are responsible for some of the remaining scatter in comparisons with simulations \citep{2015ApJ...810...21M}. The {\it shape} of the Galactic halo could potentially differentiate between different dark matter models, for example between CDM and SIDM \citep{2018MNRAS.479..359S,2018PhR...730....1T} or superfluid dark matter  \citep{2015PhRvD..91b4022K}. It can also help constrain the effect of baryons on Galactic dark matter \citep[e.g.][]{2016MNRAS.462..663B}, and at large distances can test predictions from simulations about the memory of the direction of filamentary dark matter accretion onto the halo \citep[e.g.][]{2011MNRAS.416.1377V}. Understanding the global Milky Way potential is also a necessary first ingredient to setting limits on substructure through its interactions with e.g. tidal streams (Section \ref{subsec:dmsubs}) since it is needed to construct a model of the unperturbed stream, to understand the contribution of non-regular orbits to stream structure, and to determine whether interaction rates are consistent with one dark matter theory or another (since the expected number of interactions varies with galactocentric distance; e.g. \citealt{Yoon2011,Carlberg2012}).

Many of the possible tests of dark matter in the Milky Way thus require constraints on the halo's properties on scales comparable to its virial radius, or at least to its scale length. However, the region where we have good dynamical constraints is set by where we have data: that is, where we can reliably measure distances and velocities to tracer populations. Satellite galaxies and globular clusters provide a good start, and many already have well-measured six-dimensional (6D) positions and velocities \citep[][]{2018ApJ...862...52S,2018A&A...619A.103F}, but are limited in number. It is also unclear if they represent an equilibrium population, given that in external galaxies many globular clusters appear to trace tidal features and thus could have been contributed by accreted galaxies \citep[e.g.][]{2013ApJ...768L..33V}, and given that based on cosmological simulations, we expect at least some satellite galaxies to arrive in hierarchical groups \citep[e.g.][]{2015ApJ...807...49W}. With less than 6D data, equilibrium analysis of tracers is also subject to the mass-anisotropy degeneracy.

\begin{figure*}
    \centering
    \begin{tabular}{cc}
        \includegraphics[width=0.5\textwidth]{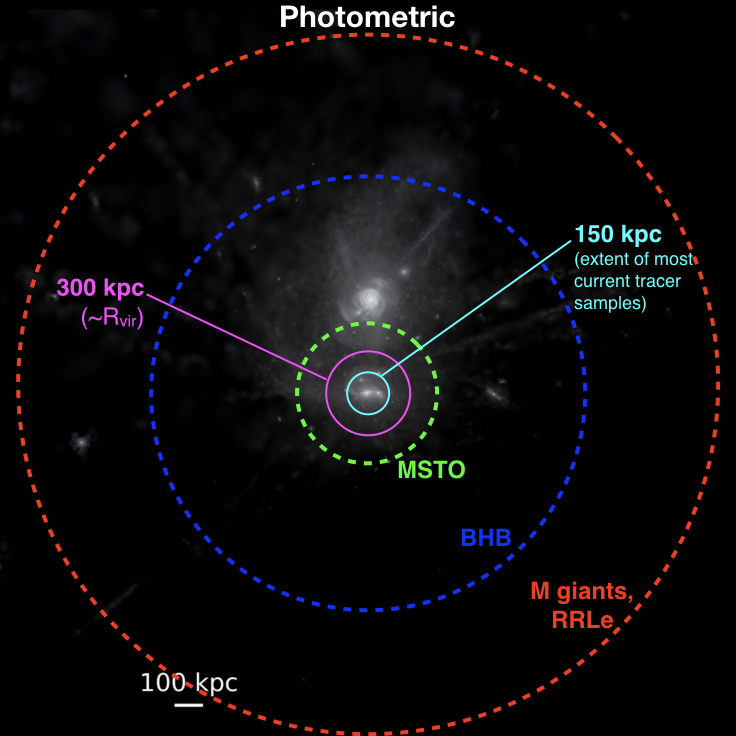} & \includegraphics[width=0.5\textwidth]{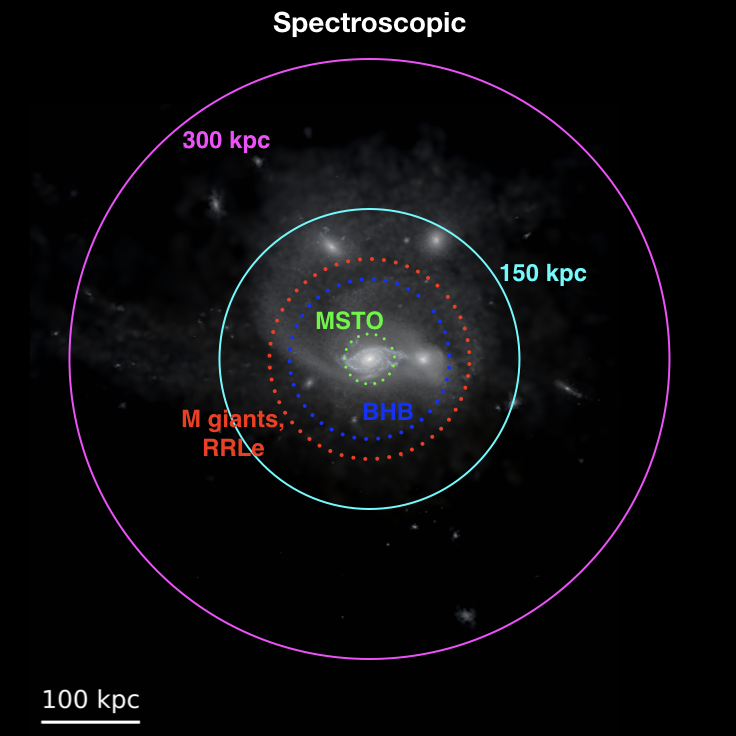}
    \end{tabular}
    \caption{{\it Left:} Distances to which LSST will detect various stellar tracers in coadded fields (limiting magnitude $r=26.7$), superimposed on an image from a cosmological-hydrodynamical simulation of a Local Group-like system \citep{2018arXiv180604143G}. {\it Right:} Distances to which planned 4-meter multiplexed spectroscopic instruments will be able to observe the same tracers, superimposed on a zoomed view of the Milky Way-like galaxy in the same simulation. To realize the promise of new, deep photometric surveys to map the Milky Way halo requires the deep, multiplexed spectroscopic capabilities of MSE.}
    \label{fig:stellar_distances}
\end{figure*}

Thus, to build an accurate map of the dark matter halo we will also need to make use of more abundant stellar tracers, and to account for the fact that at large distances these tracers are often not in dynamical equilibrium (and thus not directly suitable for, e.g., Jeans analyses). Tidal streams, which are one example of a non-equilibrium population, should be more sensitive to the halo's shape in particular (via, e.g., the precession of the orbital plane) and can extend to very large distances (the Sagittarius stream is now mapped to $\gtrsim 200$ kpc; \citealt{2017ApJ...850...96H}). The large radial range explored by a single stream can also help break degeneracies between the scale radius and total mass of the Galaxy that arise from stream modeling \citep[e.g.,][]{Bovy16a,2016ApJ...818...41S,2018ApJ...867..101B}. Gaia is providing data that will allow us to construct maps of the Galactic potential with increased accuracy using various equilibrium stellar tracers in six-dimensional phase space, to about $20 - 25$\,kpc so far \citep[e.g.,][]{2018arXiv180609635W}. For a dark matter halo of $\sim 10^{12}\,M_{\odot}$ (estimates place the Milky Way's mass in this range to within a factor of two to three), this is comparable to or a little larger than the scale radius, which is probably $1/10$ -- $1/20$ of $R_{\mathrm{vir}}$ depending on the halo concentration. Indeed, the majority of known stellar tracers in the Galaxy are strongly concentrated near its center, with only a few exceptions (Figure \ref{profile_compare}).  This is partially a function of their steep radial profile compared to what we predict for the dark matter, but also of the limiting magnitudes of current photometric and spectroscopic surveys. Crucially, while future deep photometric surveys like LSST (coadded limiting magnitude of $g=26.7$) will easily be able to identify stellar tracers all the way to $R_{\mathrm{vir}}$ (Figure \ref{fig:stellar_distances} left), MSE is the only high multiplexed spectroscopic facility under development that is capable of matching LSST's limiting photometric depth. 

Most new spectroscopic surveys with a significant stellar halo component, planned for 4-meter-class telescopes, have a limiting magnitude matched to the depth of Gaia's proper-motion survey, $g\simeq 21$ (Figure \ref{fig:stellar_distances} right), or even shallower. For comparison, to measure velocities for BHB stars ($M_g \simeq 0.5$) near the Milky Way's estimated virial radius would require spectroscopy down to $g \simeq 23$. To obtain RVs near $R_{\mathrm{vir}}$ for the extremely valuable RR Lyrae standard candles, which by the LSST era will likely be calibrated to yield distances with 2\% accuracy, will require reaching $g \sim 22 - 23$ in an integration time of $\lesssim 15$ minutes, given their typical pulsation period of $6 - 12$ hours. To reach the main sequence turnoff (MSTO), and the huge increase in tracer density that it offers, would require a depth of $g \sim 25$. Fortunately, the necessary velocity precision at these distances is not large; even at the virial radius stars should have orbital speeds of $100 - 150$\,km\,s$^{-1}$, and their motion should be primarily radial since they are mostly expected to be in tidal tails from accreted satellites, so a precision of $\sim$ 5\,km\,s$^{-1}$ is sufficient for this application. 

Besides achieving sufficient depth, the other challenge in observing distant halo tracers is their extremely low density on the sky. We expect that LSST will discover hundreds of new satellites \citep[we discuss other science with new dwarfs in Section~\ref{sec:mwlumi}, and see also][]{2014ApJ...795L..13H,Kim:2017iwr,nadler2018}, as well as tens of thousands of bright stellar tracers in distant tidal streams \citep{2017MNRAS.470.5014S}. Therefore, {\it multiplexing and multithreading} (i.e. opportunistic scheduling of individual spare fibers for distant stars in streams) will be a necessary component of these (and many other) science programs with MSE.

\subsection{Dark matter halo distortions from the LMC in the Milky Way halo}
\label{subsec:DMdist}

\begin{figure*}[b]
\centering
\includegraphics[width=6in]{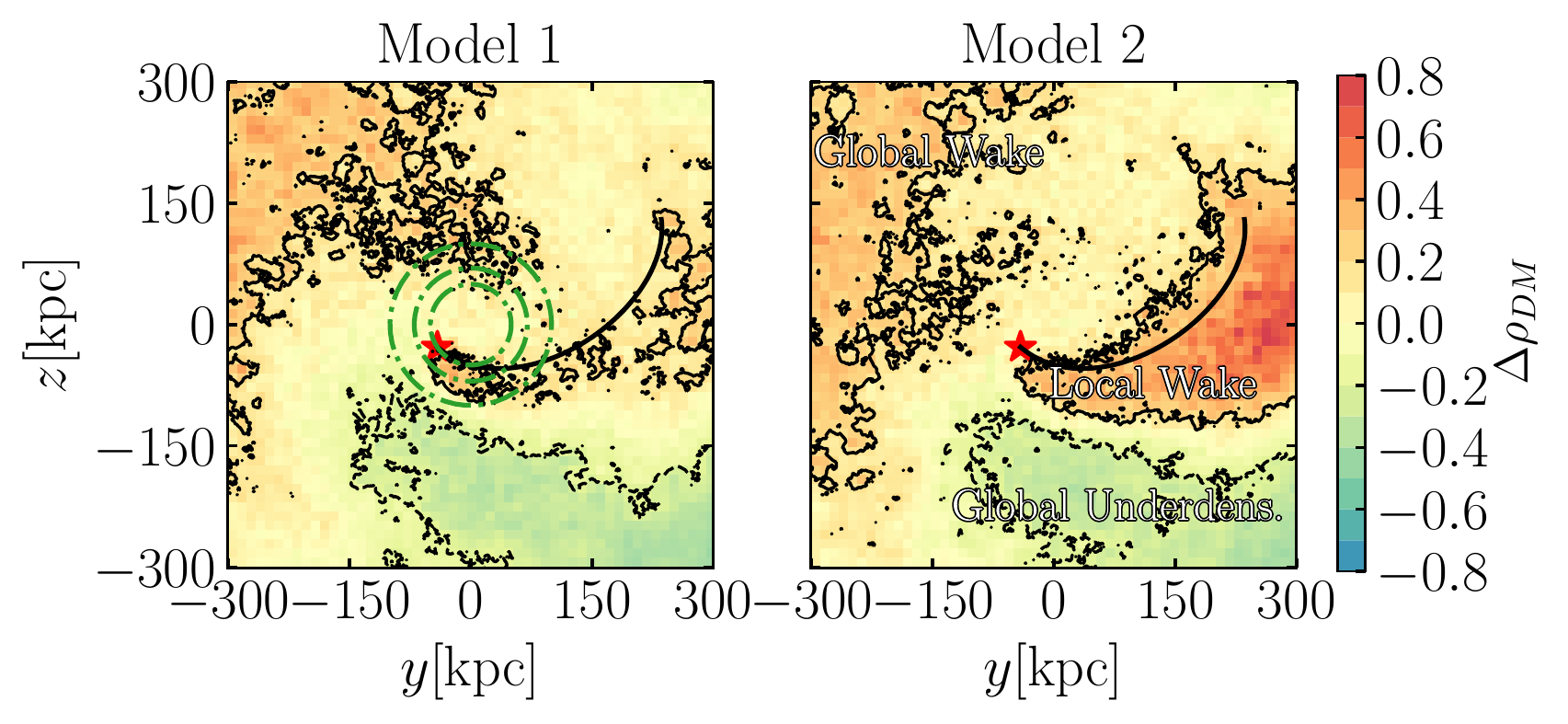}
\caption{Milky Way's dark matter density distortions, revealing the DM halo response induced by the LMC in the y-z plane (through a slab of 10 kpc thickness in the x-direction). The black line represents the LMC's past orbit. The disc is confined to the x-y plane and the Sun is at $x=-8.3$\,kpc. The green circles delineate Galactocentric distances of 45, 70 and 100 kpc. Three features are defined: 1) The local wake as the DM over-density trailing the LMC, tracing its past orbit, 2) The Global Wake as the over-density that appears in the North and 3) the Global Under-density, which are the regions that surround the Local Wake. One notices that the strength of the signal varies depending on the kinetic structure of the dark matter halo (isotropic for Model 1 and anisotropic for Model 2) which is expected due to the resonant nature of the interaction. However, the general shape of the signal is generic between models. Figure from \citet{garavito19}. }
\label{fig:garav19}
\end{figure*} 

\begin{figure*}
\centering
\includegraphics[width=6in]{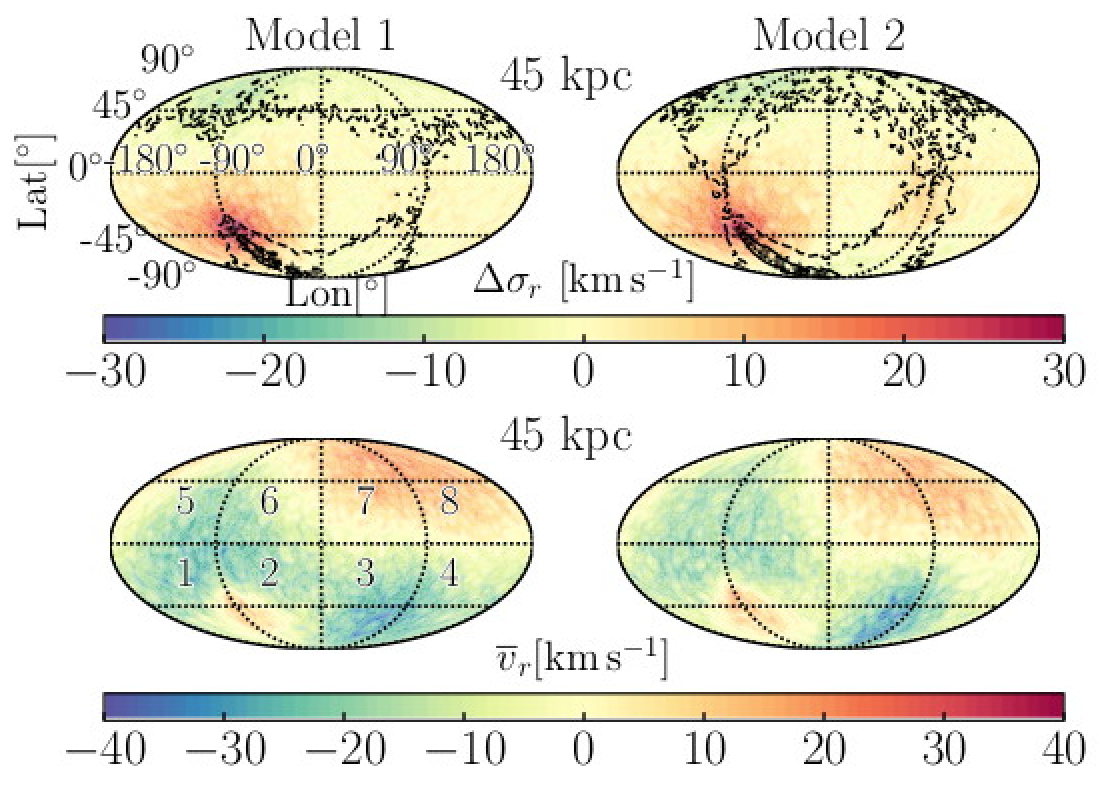}
\caption{Mollweide projections in Galactocentric coordinates of radial motions of the stellar halo induced by the LMC. Relative changes to the average 1D radial velocity dispersion are presented in the top panels, while the bottom ones show the local mean radial velocity at $r=45\,\rm{kpc}$. The contours denote the overdensities in the North and South. The past trajectory of the LMC are represented by the grey stars. $\sigma_{r}$ increases by $\sim25\,\rm{km/s}$ near the LMC in Octants 1 and 2, while in Octants 4 and 5 it decreases by $\sim14 \rm{km/s}$ to the North of the LMC, giving rise to a kinematically cold spot. The velocity dispersion in Octants 7 and 8, however remain largely unaffected. In the space of radial motions, $v_{r}$, the region in Octant 2 closest to the LMC is moving away from the Galactic Center, while the Local Wake behind the LMC in Octants 3 and 4 follows the LMC's the past orbital motion towards the Milky Way. Figure from \citet{garavito19}.}
\label{fig:radgarav19}
\end{figure*} 

The fundamental process through which galaxies merge in any dark matter model is through dynamical friction. Using linear perturbation theory in spherical systems, \cite{tremaine84, weinberg89} showed that dynamical friction is a resonant process between the orbit of a satellite galaxy and the host galaxy causing distortions of the underlying host dark matter density field from the outer to the inner region of the halo. This leads to a shift in the center of mass of the system but also higher order distortion terms which can be quantified in terms of spherical harmonics. In the Milky Way, \cite{weinberg98} used the Kalnaj matrix method \citep{kalnajs77} to demonstrated that these distortions can also lead to torques on the disc leading to warping in the case of the LMC-Milky Way interaction, for which the direct tides of the LMC are too weak otherwise to lead to a warp. This has not been appreciated in numerical N-body simulations for years leading to conflicting results, mainly due to the too poor resolution to resolve the resonant interactions at place \citep[e.g. see][for a discussion]{weinberg07}. However, more recently this has been possible due to the recent advances in computing power and increase in particle numbers able to resolve appropriately enough the phase-space structure of dark matter halos, and the effect of halo distortions on the stability of discs has been investigated in full hydrodynamical N-body cosmological settings \citep{gomez16}.

Since the recently revised proper motions of the LMC \citep{kallivayalil13}, which suggest that it is on a first infall orbit \citep{besla07}, interest in the interaction of our most massive satellite with the Milky Way has been renewed. Firstly, this has led to the possibility that the LMC may be more massive than previously thought, putting it in halos ranging from $10^{10}\,\rm{M_{\odot}}$ to $3\times10^{11}\,\rm{M_{\odot}}$.  There are various arguments suggesting for a more massive LMC, ranging from abundance matching arguments \citep[e.g.][]{moster13}, its association with the SMC and other satellites \citep[e.g.][]{D'Onghia:2008a,jethwa16,kallivayalil18} to timing argument constraints favouring a mass of $M_{LMC}=2.5^{+0.9}_{-0.8}\times10^11\,\rm{M_{\odot}}$ \citep{penarrubia16}, to name a few.

A large mass for the LMC would have dramatic consequences on the Galaxy and its subsequent modeling. In the stellar halo, this would lead to systematic biases in the modeling of tracers in the halo (globular clusters, streams, satellites, stellar halo) and as a result the inference of the orbits and associations of satellites bodies. A direct consequence of the tides from the LMC acting on stellar streams has been observed in a recent analysis of the southern part of the Orphan \citep{koposov19} showing proper motions perpendicularly offset from the stream track. \cite{erkal19} demonstrated that in order to fit the full stream a perturbation from the LMC needs to be invoked favouring a mass of $1.3^{+0.27}_{-0.24} × 10^{11} \,\rm{M_{\odot}}$. It is expected that other streams in the close vicinity of the LMC should also exhibit proper motions that are misaligned with their stream tracks, which should further help constrain the mass of the LMC \citep{Erkal2018}. 

In addition to tides, given its much larger inferred mass than typically assumed, it is expected that the whole halo of the Milky Way is also reacting. \cite{gomez15} predict a shift in the center of mass of the Milky Way due to the interaction with the Milky Way, which would result in an upward bulk motion of order $v\sim40\,\rm{km/s}$ in the stellar halo beyond a radius of $r\sim30\,\rm{kpc}$ \citep{erkal19}. Indeed, \cite{laporte18a} presented live N-body models of the interaction between the LMC and the Galaxy on a first infall orbit, showing that the resulting warp followed the lines of node of the HI warp, producing similar asymmetric distortions. This N-body experiment showed that these could result in density perturbations in the Milky Way’s dark matter halo of order 40\% around $r\sim 40 \,\rm{kpc}$ thus confirming earlier studies of the impact of halo distortions on the disc through linear perturbation theory pioneered by \cite{weinberg98,weinberg06}.

In particular, the effect of these distortions on the halo and stellar halo were studied in some depth in \cite{garavito19} where they varied the internal kinematics of the Milky Way dark matter halo and observed generic trends between realizations. Figure 
\ref{fig:garav19} shows the resulting response of Milky Way dark matter halo in those models. By generating a smooth stellar halo tracer population in the halo, they demonstrated how the signal from the halo distortions would be imprinted on the density and kinematics profiles of the stellar halo. As an example, the radial velocity signals at $r=45\,\rm{kpc}$ for the explored models are presented in figure \ref{fig:radgarav19}. For a set of reasonable assumptions on the number K-giants as a function of radius in the halo, \cite{garavito19} estimated that measuring an overdensity in the halo due to the response of the Milky Way halo to the LMC would require about $100-1000$ tracers in 20 square degree fields. Thus mapping such a signal will require a large number of tracers to probe density contrasts and bulk velocity motions making the well populated MSTO stars a prime and valuable tracer at intermediate distances in the stellar halo $r < 50 \,\rm{kpc}$. 

Measuring the wake of the dark matter halo would allow us to directly study dynamical friction in 6-D phase space in the Milky Way. Such a measurement would set strong limits on the particle nature of the dark matter (its cross-section in particular) and use the Milky Way as a complementary probe to similar analyses in galaxy clusters that examine the “wobble” of the brightest cluster galaxy (see details in Section 6.6.3). Thus MSE could provide the necessary data for three-quarters of the entire sky to map the dark matter halo’s response to the LMC \citep[e.g.][]{garavito19}.

\subsection{Identifying the dark sub-halo population with stellar streams}
\label{subsec:dmsubs}

Tidal streams are a promising tool to detect the presence of the dark subhaloes predicted by a wide range of dark matter models as summarized in Figure~\ref{interaction} \citep{ibata_etal_2002,johnston_etal_2002}. These streams form as globular clusters or dwarf galaxies are disrupted by the tidal field of the Milky Way \citep[][and references therein]{2014ApJ...795...95B}. To date, about 50 streams have been discovered in our Galaxy, with many recent discoveries aided by Gaia \citep{streams_review,des_streams,malhan_gaia_dr2,ibata2019}. Although these streams appear as coherent bands on the sky \citep[e.g.][]{fos}, they are extremely fragile and the nearby passage of a subhalo can induce relative changes to the orbits of stream stars which causes gaps and wiggles to form \citep{2008ApJ...681...40S,yoon_etal_2011,carlberg_2013,gap_evolution}. Indeed, signatures consistent with such flybys have already been claimed in the Palomar 5 \citep{carlberg_etal_pal5,stream_power,pal5_gaps} and the GD-1 stream \citep{carlberg_etal_gd1,gd1_deboer,price-whelan_bonaca_gd1}. In addition to dark matter subhaloes, baryonic substructure like giant molecular clouds \citep{amorisco_gmcs,banik2018}, the Milky Way bar \citep{pal5_gaps,2017NatAs...1..633P}, spiral arms \citep{banik2018}, as well as the disruption of the progenitor \citep{2018arXiv181107022W} can also produce perturbations in streams. Fortunately, most these effects are mitigated for streams on retrograde orbits like GD-1 \citep{banik2018, amorisco_gmcs}.

\begin{figure*}
\centering 
\includegraphics[height=6cm,trim={0.cm 0.2cm 1cm 0.5cm},clip]{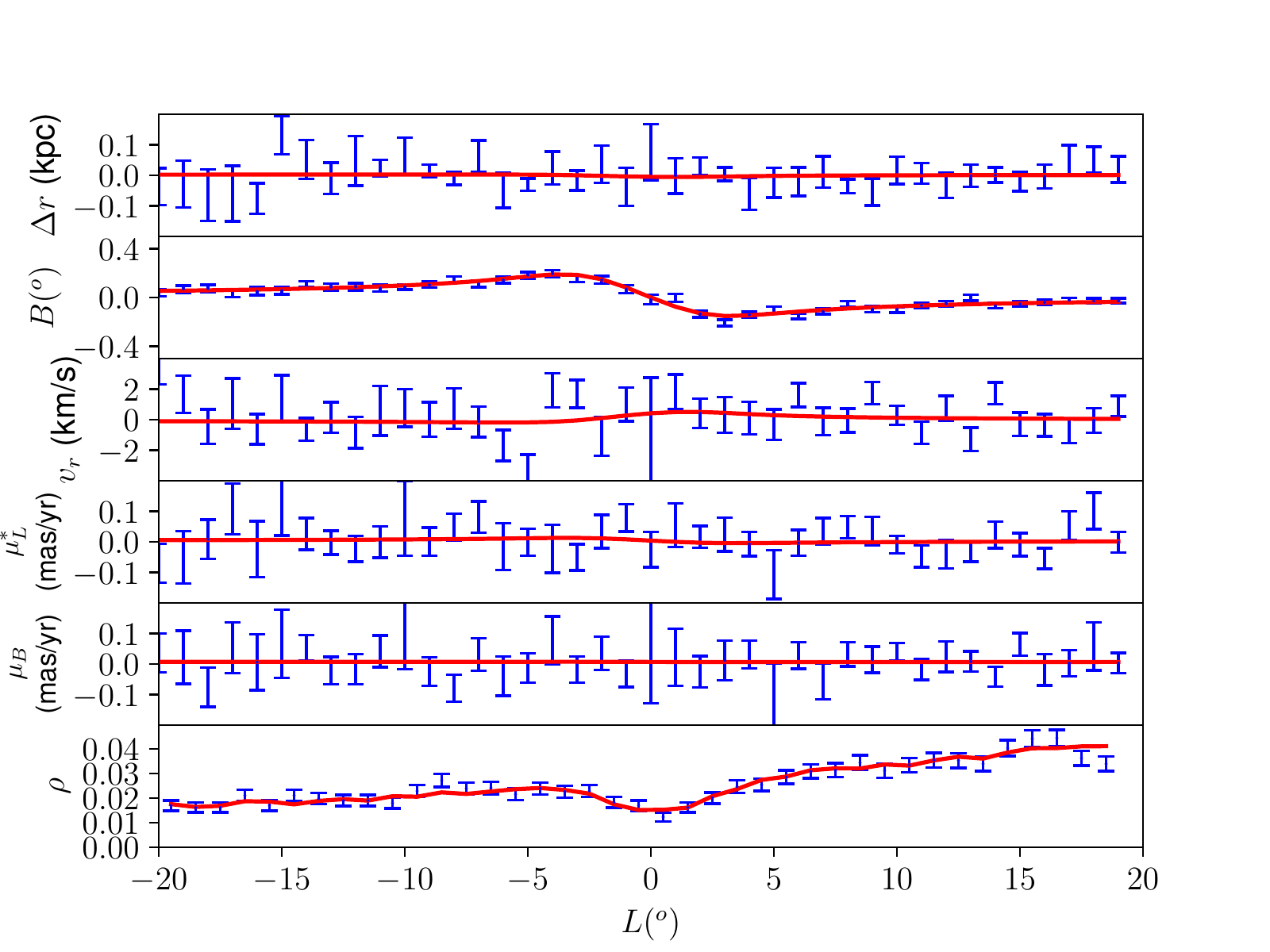}\quad
\includegraphics[height=6cm,trim={0.cm 0.2cm 1cm 0.5cm},clip]{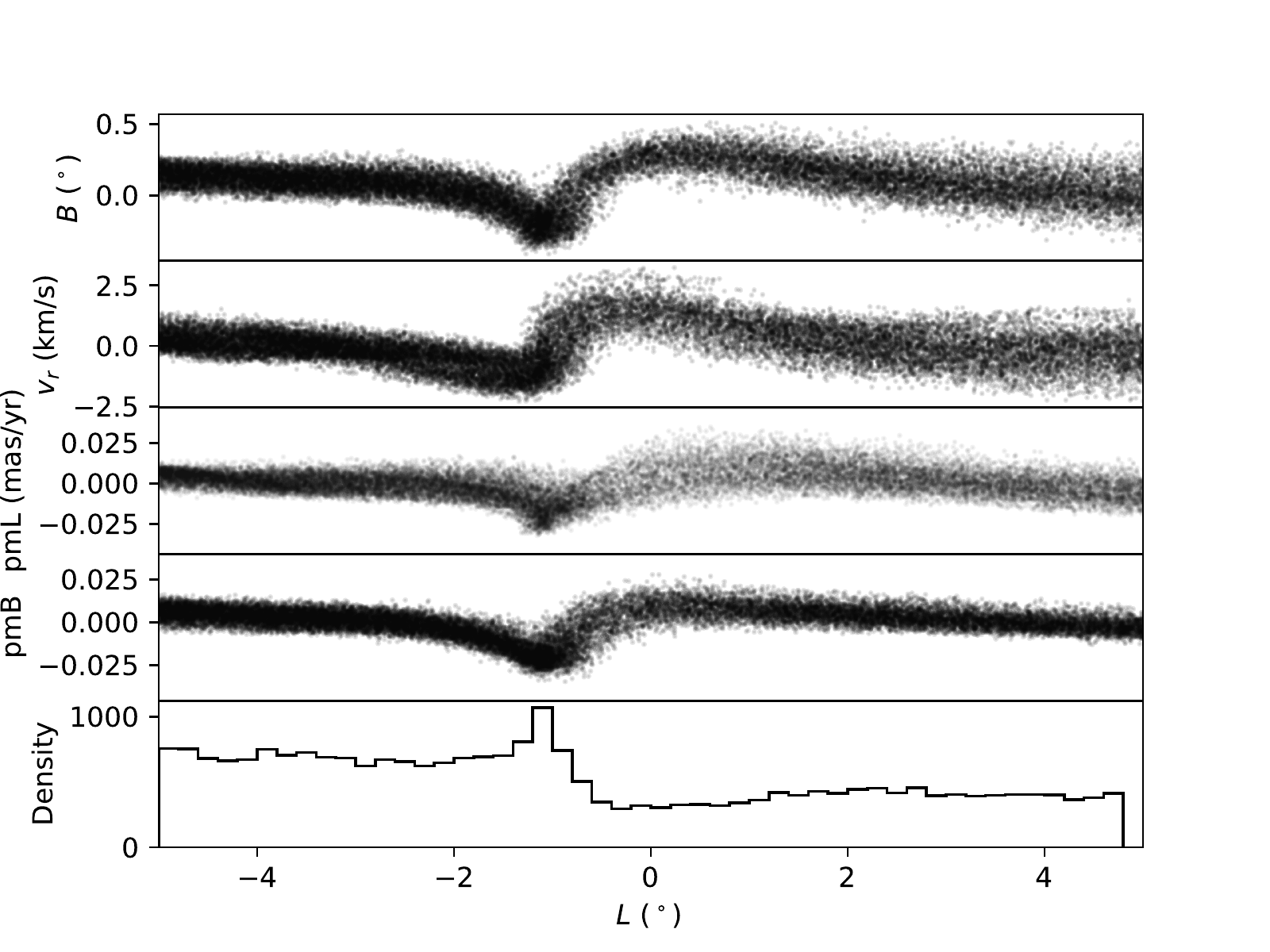}
\caption{Left: Fit of a $10^7 M_\odot$ subhalo impact adapted from \protect\cite{subhalo_properties}. The blue error bars show mock observations of an $N$-body stream impacted by a $10^7 M_\odot$ subhalo 450 Myr ago and the red line shows an analytic model which is fit to the mock data. These mock observations are made with observational errors which will be available in the near future, e.g. Gaia proper motions, radial velocities from spectroscopic surveys like WEAVE, and DES-quality photometry. Even with these errors, the fits at $10^7M_\odot$ are accurate and precise. Right: Gap in a simulated GD-1 like stream from a $10^6 M_\odot$ subhalo. This signal would be readily detectable in the density (bottom panel), on the sky (top panel), and in the radial velocities (second panel). However, the proper motions (third and fourth panel) would be undetectable even with Gaia DR2.
}
\label{fig:gaps}
\end{figure*}

Two independent techniques have been proposed for how to quantitatively extract the properties of the subhaloes which created these gaps, both of which rely on multiple dimensions of observables for each stream. First, \cite{subhalo_properties} demonstrated that each subhalo flyby produces a unique signature which can be used to almost uniquely determine the subhalo's properties, i.e. its mass, scale radius, flyby velocity, and point of impact. The inference is not quite unique since there is a one-dimensional degeneracy between the subhalo mass and its velocity relative to the stream. This degeneracy can be broken, in the first instance, by placing a prior on the subhalo's velocity, e.g. that it is bound to the Milky Way. This inference requires at least three observables of the stream, e.g. stream density, stream track, and radial velocities along the stream. This technique can be used to determine the properties of individual subhaloes and build up a catalogue of impacts. Since it determines the properties of the subhalo, it can also be used on compact baryonic substructure like giant molecular clouds. Second, \citet{stream_power} developed a statistical technique which determines the amount of substructure required to reproduce the statistical properties of the stream, e.g. the power spectrum of its density. While this technique can be used with just the stream density, it is more powerful when used with multiple observables since the gaps have correlated features in all of the observables. This technique has already been used on the Palomar 5 stream, which was found to have density variations consistent with $\Lambda$CDM \citep{stream_power}. Both of these techniques would benefit from radial velocities along the stream to faint magnitudes. 

In order to show how these gaps look in practice, Figure \ref{fig:gaps} shows two gaps produced from the nearby passage of a subhalo. The left panel shows an adapted figure from \cite{subhalo_properties} showcasing how an individual gap can be fit. The gap in this example is caused by a $10^7 M_\odot$ subhalo that impacted the stream 450 Myr ago and the properties of the subhalo can be fit up to the degeneracy described above. The right panel shows a gap in a GD-1 like stream from a $10^6 M_\odot$ subhalo. As can be seen from both panels, the signatures in the different observables are correlated which is what makes both techniques so powerful. 

\begin{figure*}
\centering 
\includegraphics[width=0.48\textwidth, trim={0.0cm 0cm 0cm 0cm},clip]{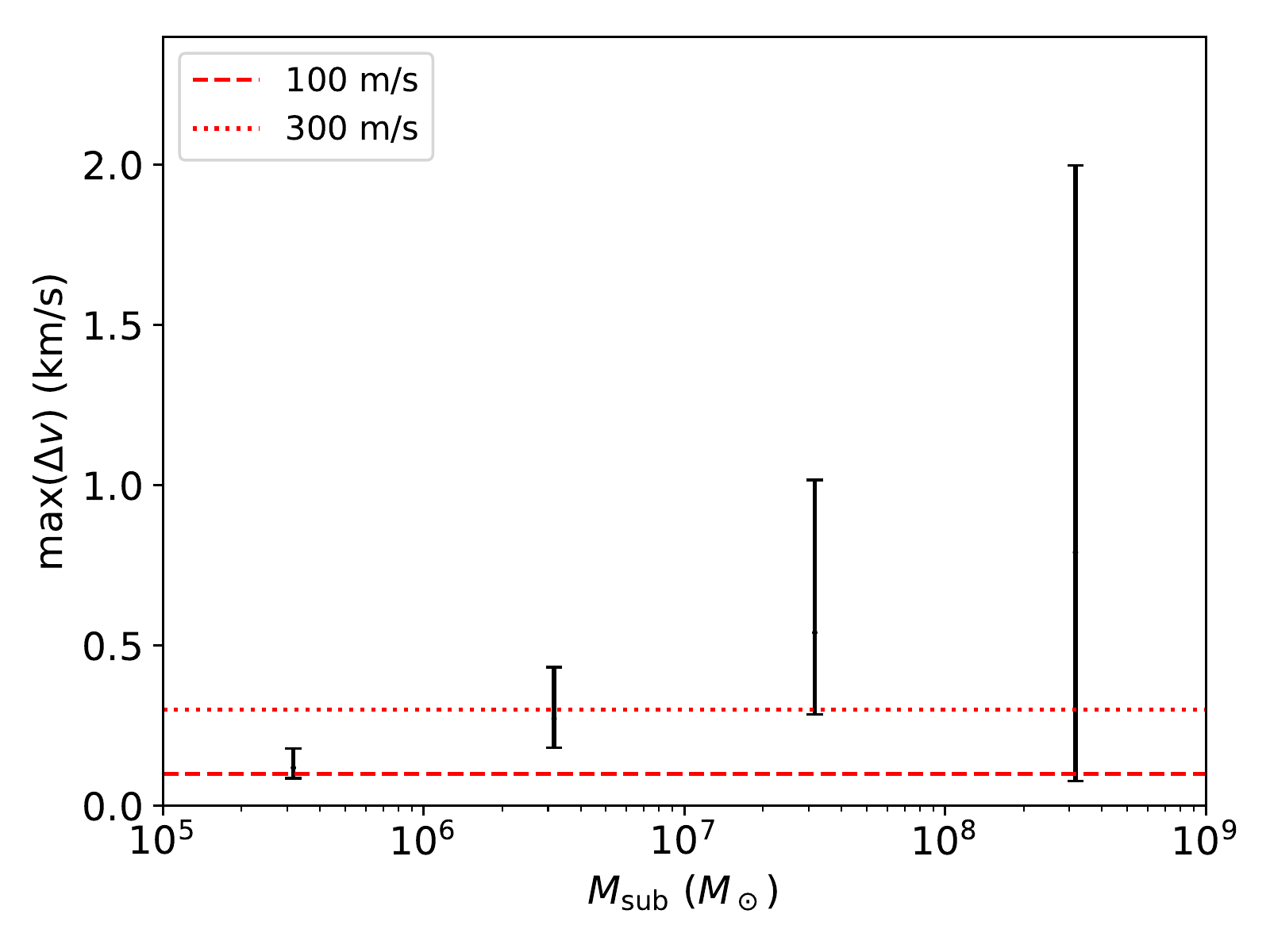}
\caption{Typical maximum velocity kick in a GD-1 like stream over a 5 Gyr duration. The black error bars show the scatter (median with 1 $- \sigma$ spread) from 1000 realizations of $\Lambda$CDM subhaloes in 4 decades of subhalo mass from $10^5-10^6 M_\odot$ up to $10^8-10^9 M_\odot$. The red dotted and dashed lines show a 300 m/s and 100 m/s uncertainty. Given an expected systematic uncertainty in the high resolution mode of $\sim 100$ m\,s$^{-1}$, MSE should be sensitive to subhaloes down to $10^5-10^7 M_\odot$. }
\label{fig:dv} 
\end{figure*}

In order to quantitatively assess what precision from MSE is needed to constrain these subhaloes, we use the results of \cite{number_of_gaps} who derived the distribution of impact properties from a distribution of subhaloes. We model the GD-1 stream which is one of the best candidates for detecting the presence of dark matter. These impact properties can then be used to determine the distribution of velocity kicks on the stream. Figure \ref{fig:dv} shows the maximum velocity kick imparted on a stream from the expected distribution of $\Lambda$CDM subhaloes over a period of 5 Gyr. This figure shows that if MSE can measure the radial velocity of stream stars down to $100 - 300$ m\,s$^{-1}$, we will be able to probe subhaloes down to $10^5-10^7 M_\odot$. Note that at the distance of GD-1, $\sim1$ km\,s$^{-1}$ would correspond to $\sim 0.02$ mas/yr in proper motion which would only be measurable for the brightest stars in Gaia DR2. Thus,  while Gaia is an excellent tool with which to measure the overall motion of the stream, spectroscopic surveys like MSE are crucial for measuring the effect of low mass substructure in velocity space. 

Finally, we note that the effectiveness of MSE will be even better than suggested in Figure~\ref{fig:dv} since the change in radial velocity occurs over a scale given by the size of the gap (see Figure~\ref{fig:gaps}). The typical size of a gap is related to the subhalo mass \citep{number_of_gaps,stream_power} with a size of a few degrees expected for a $10^6 M_\odot$ subhalo (although this gap size stretches and compresses along the orbit of the stream). Thus, instead of needing a precision of $\sim100$ m/s per star, we actually need this precision when averaging over roughly a degree. In this context, the Palomar 5 stream has $\sim100$ stars/degree brighter than $r\sim23.5$ \citep{pal5_gaps} and the GD-1 stream has $\sim10$ stars/degree brighter than $r\sim23$ \citep{gd1_deboer}. A velocity precision per star of $\sim1\,\mathrm{km\,s}^{-1}$ would allow measurements of stream perturbations at the $\sim100\,\mathrm{m\,s}^{-1}$ level for an average of 100 stars. The main limiting factor in the science described here will likely be the systematic uncertainty in MSE and not the statistical uncertainty for each star. Given an expected systematic uncertainty in the high resolution mode of $\sim 100$ m\,s$^{-1}$, MSE should be sensitive to subhaloes down to $10^5-10^6 M_\odot$. 

\subsection{Local dark matter distribution and kinematics for direct detection}

One of the possible detection mechanisms of dark matter is direct detection \citep{Goodman:1984dc}, the process in which dark matter scatters off a heavy nucleus, where the recoil of the latter emits a detectable signal. Experiments have excluded large parts of parameter space of one of the most popular dark matter scenarios for WIMPs (the most constraining limits have been performed by the Xenon1T collaboration; \citealt{Aprile:2018dbl}). The rate $R$ of this process depends on both the velocity distribution of dark matter as well as the local density of dark matter:

\begin{equation}
    R \propto \rho_{\rm{DM}} \times \int_{v_{\rm{min}}}^\infty \frac{f(v)}{v} dv, 
\end{equation}{}

\noindent where $\rho_{\rm{DM}}$ is the local dark matter density, and $f(v)$ is the local velocity distribution of dark matter. $v_{\rm{min}}$ is the minimum velocity for a particular dark matter mass that could produce a signal, and is related to the experimental threshold as 

\begin{equation}
    v_{\rm{min}} = \sqrt{\frac{Q m_N}{2 \mu^2}}, 
\end{equation}

\noindent where $Q$ is the recoil energy, $m_N$ the mass of the nucleus (Xenon for example), and $\mu = m_\chi m_N / (m_\chi + m_N)$ is the reduced mass of the dark matter $m_\chi$ and the heavy nucleus. 

Astrophysical errors on the local density of dark matter can change current limits from $\sim 30\%$ to a factor of two, depending on the method used (see \citealt{Read:2014qva} for a review). Using local stars as tracers, it is possible to reduce the errors on the local measurement of dark matter local density  \citep{2012ApJ...756...89B,2014MNRAS.445.3133P}. A particular systematic in the measurement of the dark matter density is the uncertainty in the density and distribution of the baryonic component (stars and gas). With radial velocities from MSE for a large set of stars, we will be able to improve on the existing measurements. Coupled with the proper motions of Gaia, we expect to resolve smaller structures and improve our understanding of the baryonic components, leading to a more accurate measurement of the local density of dark matter.

Another potential reducible systematic is the local velocity distribution. A new strategy to empirically obtain the velocity distribution of dark matter from metal poor stars has been introduced in \cite{Herzog-Arbeitman:2017fte} for the case of the metal-poor relaxed component, and in \cite{Necib:2018b} for the case of more recent mergers. These metal poor stars have mostly been accreted, like dark matter, and hence the most metal poor stars should trace the velocity distribution of the oldest dark matter component. Such a correlation has been used to determine the local velocity distribution from Gaia in \cite{Herzog-Arbeitman:2017zbm} and \cite{necib2018}. Here,  a new structure in velocity space called Gaia Enceleadus \citep{2018MNRAS.477.1472B,2018MNRAS.478..611B,2018Natur.563...85H,2018MNRAS.475.1537M} has been modeled for its dark matter content. 

In order to get the most detailed velocity distribution of dark matter locally, an accurate measurement of the metallicity as well as the 3D velocities of a large number of nearby stars is required. Gaia DR2 has already shown the capability of finding nearby velocity substructure \citep{2018MNRAS.475.1537M}. MSE will be able to provide the missing radial velocity component of all Gaia stars across its full magnitude range, and will extend even further to 23rd magnitude to match with future space mission such as WFIRST. At the bright end, the estimated errors on the radial velocities from MSE stars will be of order hundreds of m\,s$^{-1}$ or better. When coupled with excellent distance measurements from Gaia, this will provide the best set of 3D velocity measurements that exist. The dominating errors in that case will be systematics of the strength of the correlation between the dark matter and the stars \citep[e.g.,][]{2018arXiv181111763B}.

\subsection{Dark matter distribution in the Galactic Center for indirect detection}\label{s:indirect}

Dark matter could also be detected indirectly through its self-annihilation or decay into Standard Model particles, such as gamma rays, neutrinos, electrons and positrons. The likely sources for this search are places known to have a dense concentration of dark matter, including the Galactic Center and the satellites of the Milky Way. 
The self-annihilation or decay processes depend on the density of dark matter at the source, and the rate is therefore condensed into a parameter called the J-factor, defined as

\begin{equation}\label{eq:jfactor}
    J = \int_{\rm{l.o.s}} (\rho(l))^p dl,
\end{equation}

\noindent where $\rho$ is the density of dark matter, $p$ is the number of dark matter particles participating in the interaction, $p=1$ for decay and $p=2$ for annihilation, and the integral is set along the line of sight from the object to the experiment. 

Getting accurate measurement of the velocities of stars closer to the center is crucial in obtaining the correct density profile of the dark matter. Although the field at the Galactic center is crowded, MSE will be able to get radial velocity measurements for a large number of stars (with better than $10$\,km\,s$^{-1}$ precision) within a few kiloparsecs of the Galactic Center.
We can couple radial velocity measurements of MSE with Gaia to find the escape velocity at different Galactocentric distances (similarly to the analysis in \citealt{2018A&A...616L...9M}), or work using the radial velocity alone but explore a larger range of distances surpassing Gaia measurements (e.g., \citealt{2017MNRAS.468.2359W}).
This leads to a better determination of the slope of the density of dark matter, and therefore would constrain the J-factor for indirect detection. 

\section{Dwarf galaxies in the Milky Way and beyond with resolved stars}\label{sec:dwarfs}

Galaxies like the Milky Way are expected to contain a plethora of dark matter subhaloes \citep{klypin_etal_1999}. 
The largest of these subhaloes, above $\sim10^8 M_\odot$, are large enough to have hosted star formation in the early Universe and are thus visible as satellite galaxies \citep{Jethwa:2016gra,Kim:2017iwr,norman2018,wheeler2018}. 

Local Group dwarf galaxies are attractive targets for investigating the nature of dark matter due to their proximity, large dynamical mass-to-light ratios, and early formation times.
The standard cosmology model predicts the abundance and internal structure of the dark matter halos that host dwarf galaxies.
The particle physics governing dark matter could lead to observable consequences including reducing the number of dwarf galaxies, flattening the density profiles of their dark matter halos, or producing energetic Standard Model particles through annihilation or decay. 
MSE's ability to gather large stellar-kinematic samples for faint dwarf galaxies, combined with data from X-ray and gamma-ray observatories, will be crucial for testing these predictions and illuminating the physical nature of dark matter.

The theoretical landscape of dark matter models described in Section~\ref{sec:justification} clearly provides support for searches of deviations from the cold collisionless dark matter idea, using Local Group dwarf galaxies. 
Several apparent discrepancies between CDM-simulated and the observed Universe, particularly on the smallest galactic scales, also motivate such searches. 

Around galaxies with $\mathrm{M_{vir}} \sim 10^{12}$M$_\odot$, cosmological N-body simulations that consider only gravitational interactions among CDM particles typically form $\sim 10$ times more subhalos with $\mathrm{M_{vir}} \sim 10^8$M$_\odot$ than have been detected as luminous dwarf-galactic satellites of either the Milky Way or M31. This has been dubbed the ``missing satellites'' problem~\citep{klypin_etal_1999, Moore1999}. However, the ultrafaint dwarfs discovered recently seem to bring the census of dwarf galaxies into agreement with predictions of the CDM model, after taking into account detectability and the impact of reionization~\citep{2018PhRvL.121u1302K}.

\begin{figure*}
\centering
\includegraphics[width=4in]{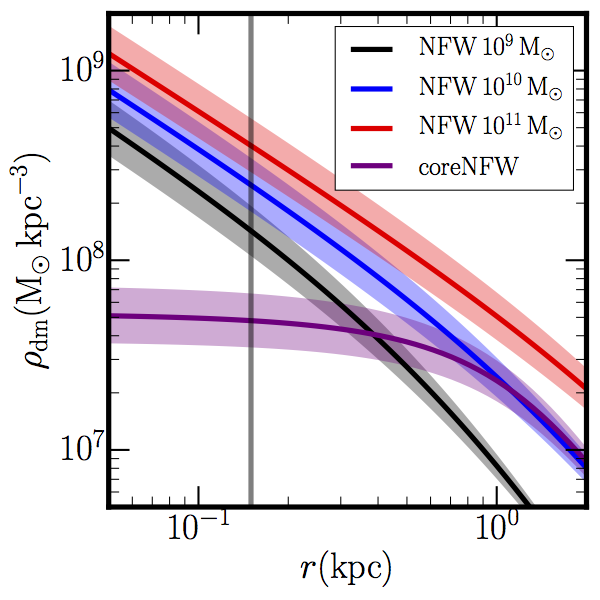}
\caption{Density profiles for dark matter halos having virial masses in the range expected for dwarf galaxies.  Black,
blue and red curves correspond to the pure `NFW' profiles that characterize standard cold dark matter halos.  The purple curve shows how the halo with $M_{200}\sim 10^{10}M_{\odot}$ might evolve in response to energetic feedback from star formation, lowering its central density and flattening the inner density gradient.  Figure from \citet{Read2018a}.
\label{f:read18}}
\end{figure*} 

The dynamical masses of observed satellites (estimated within their half-light radii) are systematically smaller than the masses (evaluated at the same radii) of halos in simulations \citep{Boylan-Kolchin2011, Boylan-Kolchin2012}.  This ``too big to fail'' problem is perhaps a symptom of a further discrepancy between the shapes of simulated and observationally-inferred mass-density profiles, $\rho (r)$ -- the ``cusp-core" problem alluded to previously. This discrepancy could be pointing to a deviation from the CDM paradigm, but it could also be correlated with ``baryonic physics" including feedback from star formation, as discussed in Section \ref{sec:justification}. 

Current N-body+hydro simulations agree with simple expectations 
that feedback processes become inefficient in the least luminous, most dark-matter-dominated galaxies, such that dwarf galaxies with $L < 10^6$L$_\odot$ should retain their primordial CDM cusps \citep[][ Figure \ref{f:read18}]{Penarrubia2012, Garrison-Kimmel2013}.  This implies that dark matter physics can be separated from the astrophysics of galaxy formation, provided that observations can constrain the density profile of ultrafaint dwarfs. 

The current census of Local Group dwarf galaxies includes $\sim 50$ such systems, $\sim$10 of which host more than $10^3$ stars brighter than $V \sim 23$.  For these galaxies, an MSE survey will deliver stellar-kinematic samples as large as those currently used to distinguish dark matter cores from cusps in dwarf galaxies with $L > 10^6$L$_\odot$ \citep{Walker:2011, Read2018a, Read2018b}.  Thus, MSE will have unprecedented power to constrain the nature of dark matter by measuring the density profiles of dwarf galaxies.  

\subsection{Luminosity function of Milky Way satellites in the era of LSST}\label{sec:mwlumi}

If the dark matter distribution is well-described by numerical simulations of cold, collisionless dark matter, LSST is expected to discover another $\sim 200$ dwarf galaxies out to the virial radius of the Milky Way \citep{Tollerud:2008ze,Hargis:2014kaa,Kim:2017iwr,newton2018,nadler2018,kelley2018}.  Spectroscopic followup of these objects will require a large amount of telescope time on a 10+ meter class telescope.  Such observations are necessary to provide stellar chemo-dynamic samples with sufficient precision ($\sim 1$ km s$^{-1}$ velocities, $\sim 0.1$ dex metallicities) to distinguish dark-matter-dominated dwarf galaxies from outer halo star clusters, to estimate dynamical masses and metallicity distributions, and to measure systemic line-of-sight velocities. 

\begin{figure*}
\centering
\includegraphics[width=4in, trim=0.25in 2.in 0.5in 0.5in,clip]{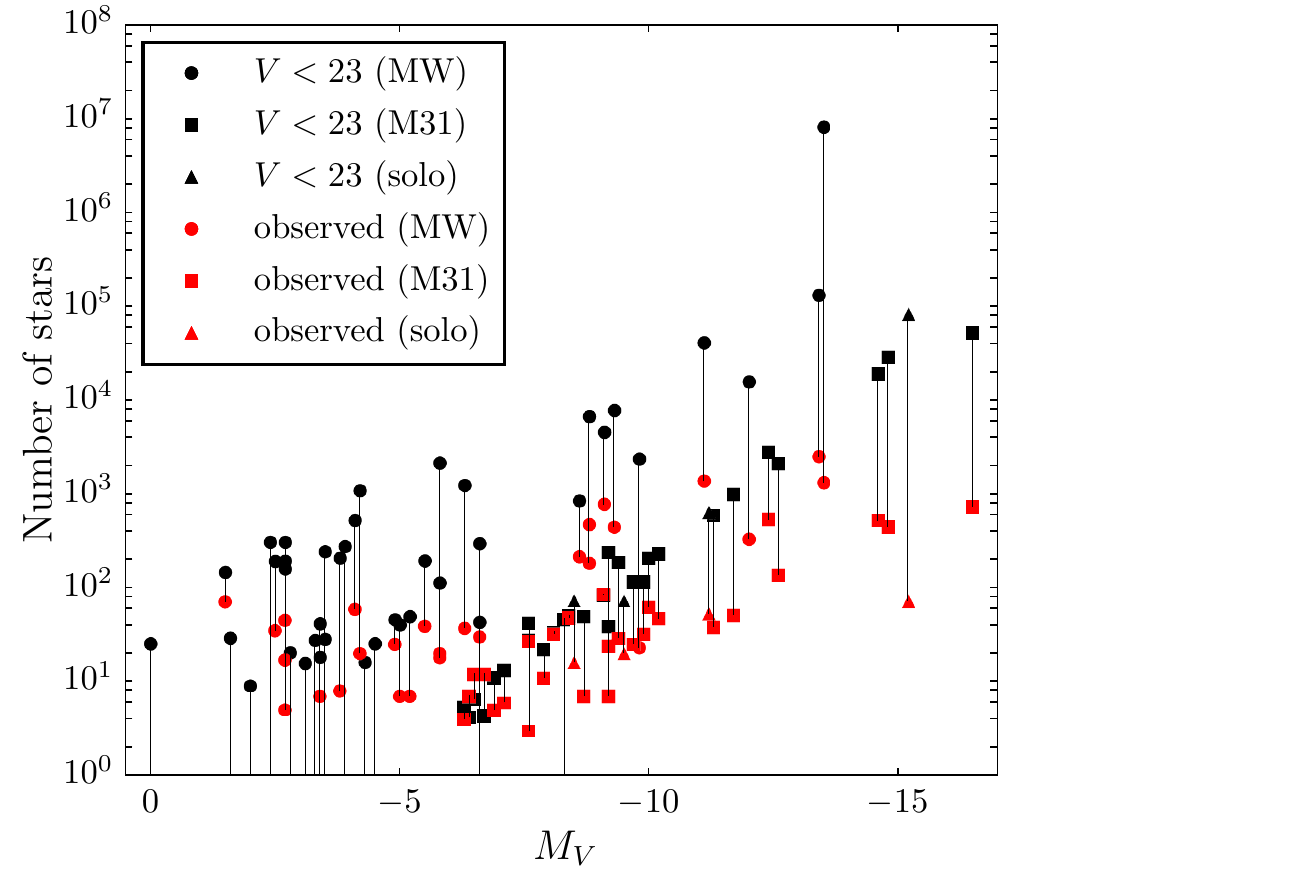}
\caption{Spectroscopic sample sizes for known Local Group dwarf galaxies with $M_V\gtrsim -16.5$.  Black points indicate number of stars stars brighter than a fiducial magnitude limit of $V<23$ that would be observable with MSE.  Connected red points indicate the current sample size available in the literature.  Marker types specify whether the dwarf galaxy is a satellite of the Milky Way (circles), a satellite of M31 (squares), or an isolated system (triangles). Note that the sample size is displayed on a logarithmic scale. 
\label{f:dwarfs_numbers}}
\end{figure*}

In order to quantify the contributions that MSE can make towards understanding the dark matter content of the Galactic satellite population, we first consider the spectroscopic sample sizes that would be achievable in an MSE survey of Local Group galaxies.  For \textit{known} systems less luminous than M32 ($M_V\gtrsim 16.5$), we estimate the number of stars brighter than a fiducial magnitude limit of $V\leq 23$ by integrating a log-normal stellar luminosity function \citet[][]{Dotter2008} of a 10-Gyr-old stellar population with the metallicity reported by \citet{McConnachie12}.  Figure \ref{f:dwarfs_numbers} compares these numbers to the largest spectroscopic sample sizes that are currently available in the literature for each observed galaxy.  In most cases, an MSE survey would increase the available spectroscopic sample by more than an order of magnitude.  For the `ultrafaints' with $M_V\gtrsim -5$, this would imply stellar samples of several hundred to a thousand member stars.  For the Milky Way's `classical' dwarf spheroidals ($-7 \gtrsim M_V\gtrsim -13$) and more luminous distant objects (e.g.,  NGC 185, NGC 205 and NGC 6822), it means samples reaching into the tens of thousands of member stars.  For each class of object, the precision with which we can infer dark matter densities from line-of-sight velocities alone increases by more than an order of magnitude.  Used in combination with proper motion data,  from the final Gaia release, 30m-class telescopes, or future space missions (e.g., WFIRST, Theia),  MSE data will provide definitive constraints on the inner density profiles that distinguish various particle physics models.  

\begin{figure*}
\centering
\includegraphics{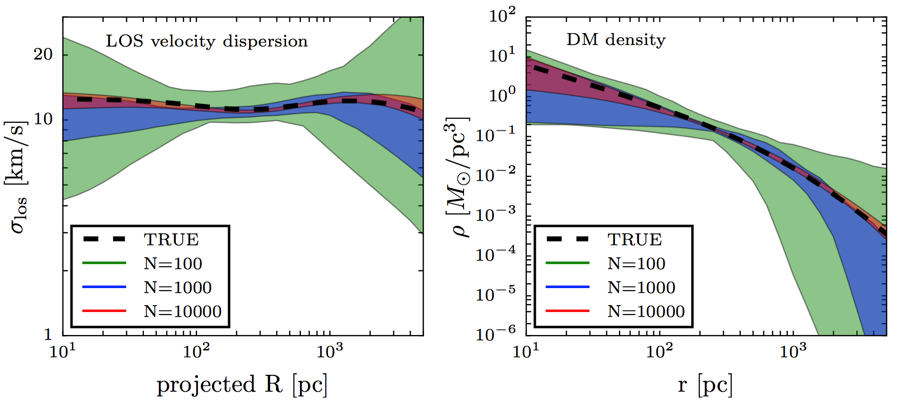}
\caption{Recovery of intrinsic line-of-sight velocity dispersion (left) and inferred dark matter density (right) profiles as a function of spectroscopic sample size. Shaded regions represent 95\% credible intervals from a standard analysis (based on the Jeans equation) of mock data sets consisting of line of sight (LOS) velocities for $N = 10^2$, $10^3$ and $10^4$ stars (median velocity error 2\,km\,$^{-1}$), generated from an equilibrium dynamical model for which true profiles are known (thick black lines, which correspond to a model having a cuspy NFW halo with   $\rho(r)\propto r^{-1}$ at small radii).
}\label{f:dm_profile}
\end{figure*} 

To gauge the effects of stellar sample size on inferences about dark matter, we compare results from dynamical analyses of three mock data sets consisting of N = $10^2$, N = $10^3$ and N = $10^4$ stars. In each case, the artificial sample is drawn from a phase-space distribution function describing a stellar population that follows a Plummer surface brightness profile and traces a gravitational potential dominated by a NFW dark matter halo. The analysis uses standard Bayesian procedures to fit simultaneously for the velocity anisotropy, surface brightness and dark matter density profiles (a by-product is specification of the line-of-sight velocity dispersion profile), similar to the procedures described by \citet{GeringerSameth2015, Bonnivard2015}.

Figure \ref{f:dm_profile} compares the resulting inferences for velocity dispersion and dark matter density profiles, displaying bands that enclose 95\% credible intervals in each case. Given the samples that MSE can provide, inferences about kinematics and dark matter content of dwarf galaxies will become dramatically more precise.  We will infer the dark matter densities of ultra-faint satellites with precision similar to what is achieved today only for the most luminous classical dwarfs, for which we will infer density profiles with unprecedented precision from pc to kpc scales. This improvement will render dynamical analyses limited by systematics (e.g., triaxiality, non- equilibrium kinematics, unresolved binary stars) instead of statistics.  Moreover, since the smallest ultra-faints have half-light radii of just a few tens of parsecs, large MSE samples for these objects will provide strong constraints on dark matter densities at these smallest galactic scales. 

\subsection{Precise determination of the J-factor of nearby ultra-faint dwarf galaxies}\label{sec:gammarays}

Due to the high dark matter densities and the lack of astrophysical backgrounds (e.g., pulsars, scattering of cosmic rays off ISM, etc.) that contaminate searches near the Galactic center (Section~\ref{s:indirect}), dwarf galaxies represent the cleanest available targets in searches for annihilation and decay signals \citep{gunn78, Lake1990}.  

\begin{figure}
\centering
 \begin{tabular}{ll}
 \hspace{-0.1in}\includegraphics[width=0.52\textwidth]{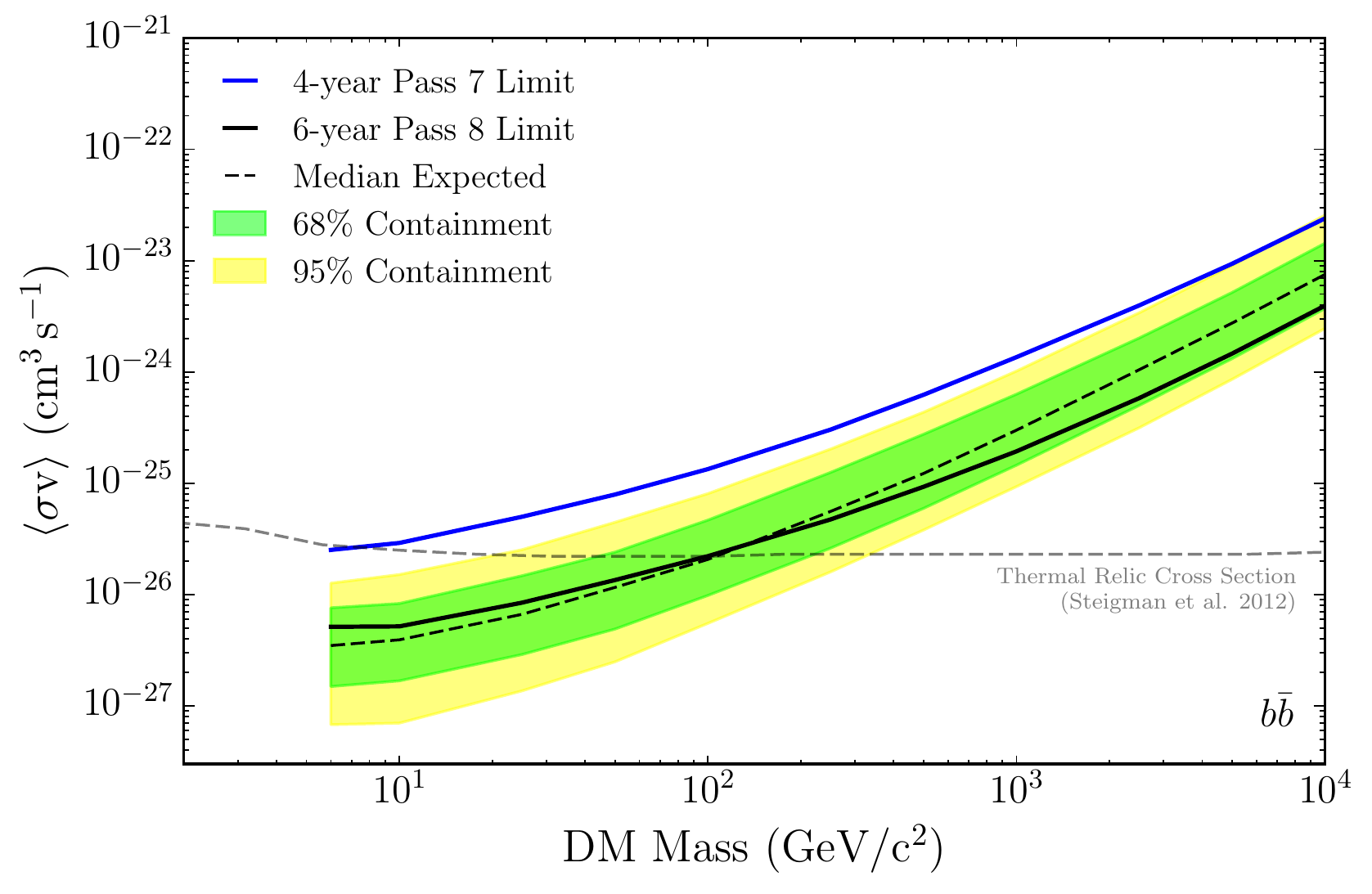}&\hspace{-0.1in}\includegraphics[width=0.52\textwidth]{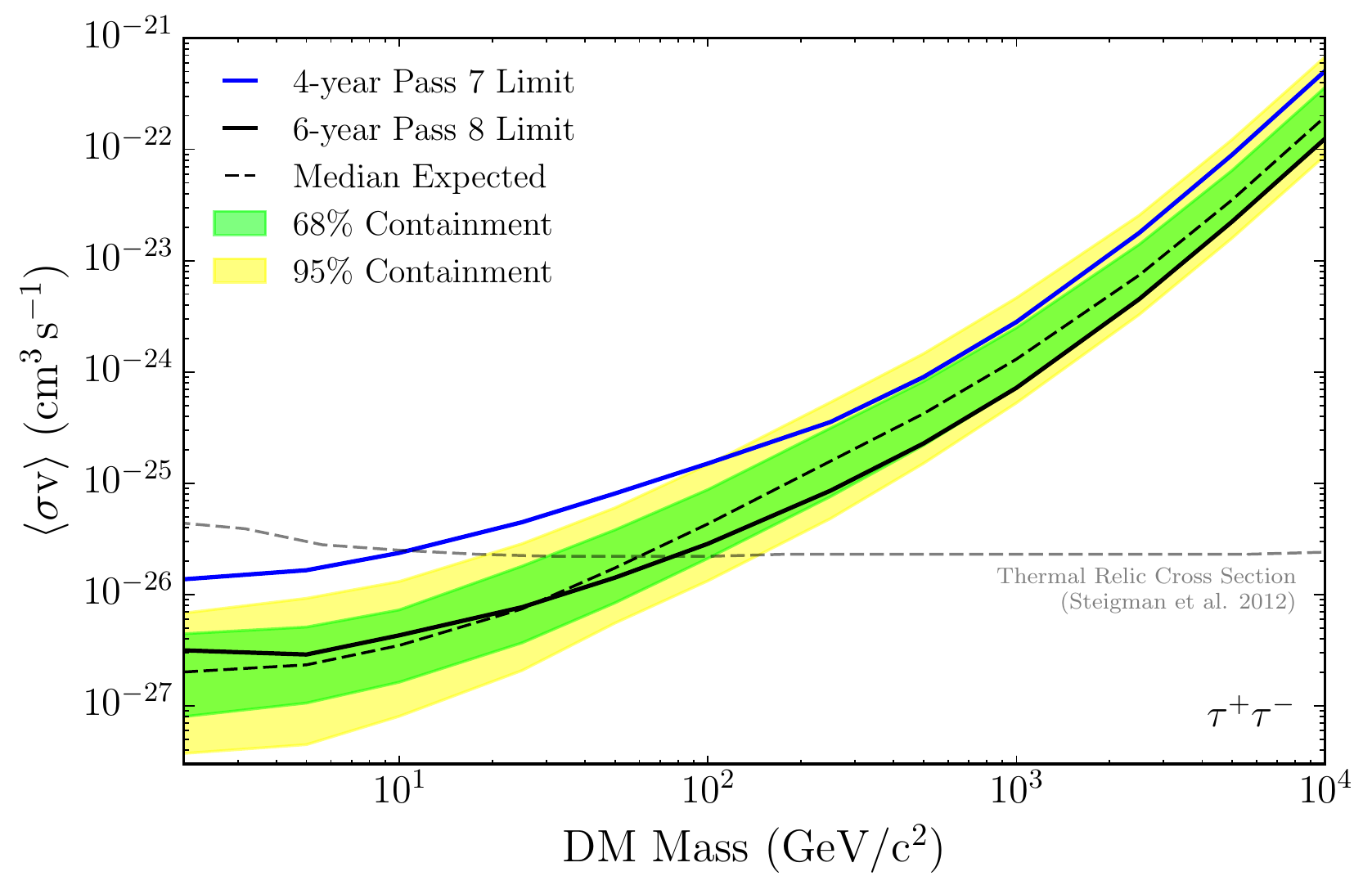}\\
 \end{tabular}
  \caption{Upper limits on the cross section for dark matter annihilation into final states of quarks ({\it left}) and leptons ({\it right}), derived from analysis of gamma-ray data and stellar spectroscopy of 15 Galactic satellites.  Dashed lines show the median expected sensitivity, while bands indicate 68\% and 95\% quantiles.  Dashed gray lines indicate the cross section expected for a thermally-produced weakly interacting massive particle \citep{Steigman:2012nb}. Figure from \citet{Ackermann2015}.}
  \label{f:fermi}
\end{figure}

The flux of photons received from annihilation and decay of dark matter is proportional to the $J$-factor given by Equation~\ref{eq:jfactor}.  Thus, given a measurement of photon flux, or even a non-detection, one can use stellar-kinematic estimates of the dark matter density profile to infer or constrain relevant particle physics properties.  For example, Figure \ref{f:fermi} shows constraints on the dark matter self-annihilation cross section as a function of particle mass \citep{Ackermann2015, GeringerSameth2015}.  These upper limits are derived by combining non-detections of gamma-rays from the Fermi-LAT with density profiles estimated from stellar-kinematics of fifteen of the Milky Way's dwarf satellites.  For particle masses $M_{\chi} < 100$ GeV, these limits begin to constrain the cross section that would naturally give the cosmologically-required $\Omega_\mathrm{DM} \sim 0.2$ in the case of thermally-produced WIMPs \citep{Steigman:2012nb}.  Stellar-kinematic data have also been used to evaluate the significance of reported decay signals in X-ray observations of individual dwarf spheroidals \citep{Loewenstein2010, Boyarsky2010, Jeltema2015, Ruchayskiy2016}.

Regardless of whether an unambiguous photon signal is ultimately detected, the resulting inferences about particle properties are only as good as estimates of dark matter densities (via the $J$-factor) derived from stellar kinematics. As shown in \citet{Albert2017} decreasing the uncertainty in J-factor from 0.6 dex to 0.2 dex, can result a factor of $2 - 3$ improvement in the sensitivity of constraints on the annihilation cross-section. This dependence highlights the impact MSE will have on efforts to determine dark matter's particle nature.  Among the known dwarf galaxies, the most attractive targets for annihilation/decay searches are also the least luminous ($L_V < 10^3$L$_\odot$), primarily because these happen also to be the nearest.  However, while the dynamical masses of these ultrafaint systems are also consistent with extremely large dark matter densities \citep{Martin:2007ic, sg07, martinez11, simon11}, 
the small number of spectroscopic measurements and the possibility of binary orbital motion inflation the velocity dispersion~\citep{McConnachie2010,minor10} have been serious hurdles.  
Improving upon existing samples by more than an order of magnitude, MSE will have a major impact in this area, which represents one of the best opportunities to resolve dark matter's particle nature. 

\subsection{Controlling systematics with spatial and temporal completeness at high resolution}

Nearly all methods for inferring the amount and distribution of dark matter within dwarf galaxies rest on the assumption that the observed stellar kinematics faithfully trace the underlying gravitational potential.  However, the degree to which tidal effects and orbital velocity of binary stars invalidate this assumption are poorly constrained.  Moreover, the ability to resolve the internal velocity dispersion $\lesssim 4$\,km\,s$^{-1}$ of the ultra-faint galaxies is limited by the spectral resolution of existing instrumentation.  MSE will solve these problems by observing thousands of targets simultaneously over a wide field at a  resolution capable of measuring sub-km\,s$^{-1}$ dispersion, enabling inexpensive repeat observations.  As a result, MSE will deliver spectroscopic samples with unprecedented completeness in both the spatial and time domains at high resolution.  These capabilities are crucial for obtaining accurate dynamical masses for the nearest ultrafaint dwarf galaxies, which are the most important targets for indirect detection searches.

Nearly all published spectroscopic data sets for Galactic satellites suffer from spatial incompleteness; more specifically, selection biases that leave outer regions either under-sampled or neglected altogether. The outer regions of dwarf galaxies are important for investigating the outer structure of dark matter halos \citep{Walker2007}, identifying kinematic signatures of tidal disruption \citep{Munoz2006}, mapping the evolution of internal halo structure \citep{El-Badry2016}, measuring systemic proper motions \citep{Kaplinghat2008, Walker:2008ax}, and tracing stellar population gradients \citep{Harbeck2001, McConnachie2007}.
To date, the relatively low fractions of bona fide members at large radii has made thorough observations of these regions prohibitively expensive.

Existing spectroscopic samples also suffer from \textit{temporal} incompleteness.  For the vast majority of measured stars, published velocities are based on a single observation.  However, in cases where repeat observations exist, there is evidence for velocity variability of individual stars \citep{Olszewski1996, minor10, simon11, koposov11, Spencer2017}, most likely due to the internal motions of unresolved binary systems.  Radial velocity surveys of the Galactic halo indicate that stellar multiplicity increases toward lower metallicity, suggesting that the binary fractions in dwarf galaxies may be large.  Indeed, recent studies of the limited multi-epoch data sets available for dwarf galaxies estimate binary fractions in the range of $50-75\%$ \citep{minor13,Spencer2018}.  Since binary motions alone can contribute velocity dispersions of $\sim 2-3$ km s$^{-1}$ \citep{McConnachie2010,minor10}, such contamination can potentially dominate the dynamical masses estimated for the coldest ultra-faint systems, which typically have intrinsic dispersions estimated to be $\lesssim 3$ km s$^{-1}$ \citep{Martin:2007ic, sg07, Caldwell2017}.  The possible inflation of dynamical mass measurements of ultra-faint dwarfs presents a major caveat for conclusions about dark matter physics that rely on ultra-faint dwarfs, including constraints on indirect detections with gamma-rays (Section~\ref{sec:gammarays}).

\begin{figure*}
\centering
\begin{tabular}{ll}
\includegraphics[width=6in,trim=0in 3.25in 0in 0in,clip]{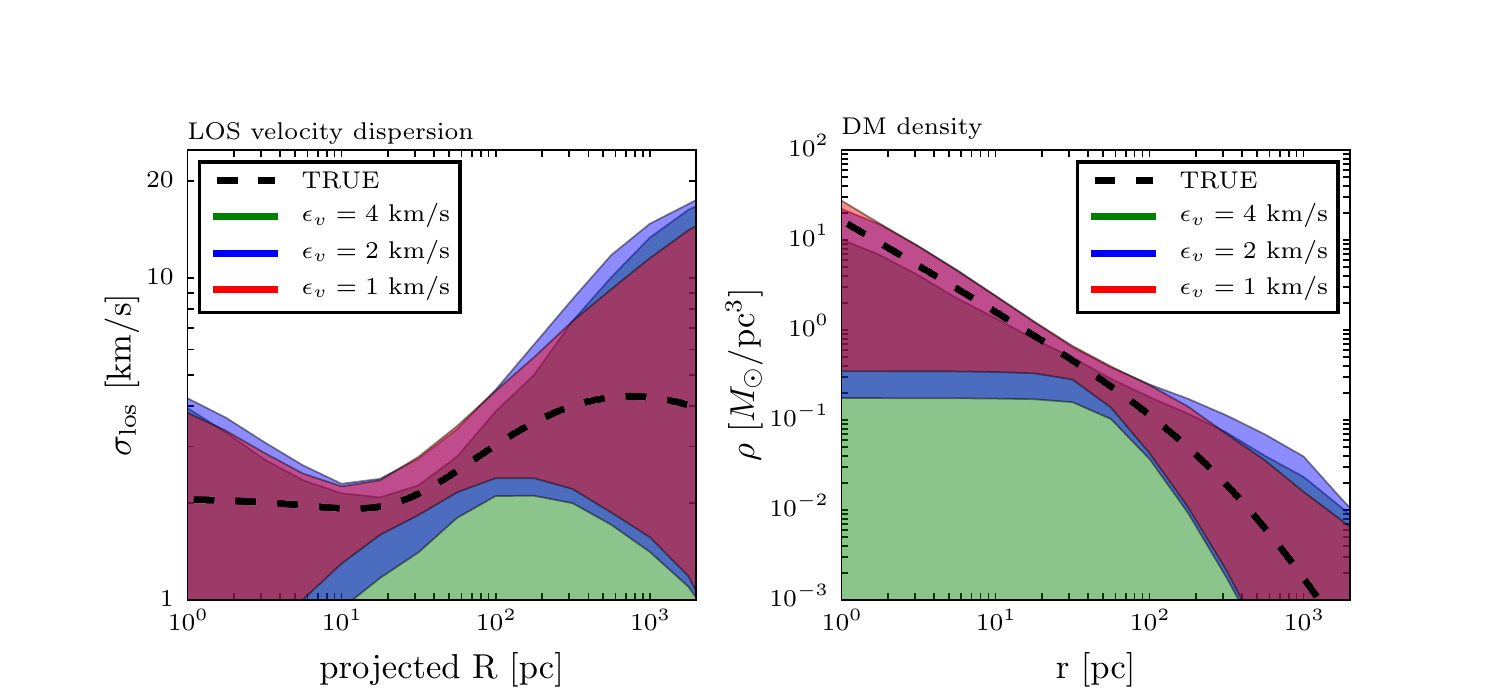}\\
\includegraphics[width=6in,trim=0in 3.25in 0in 0.2in,clip]{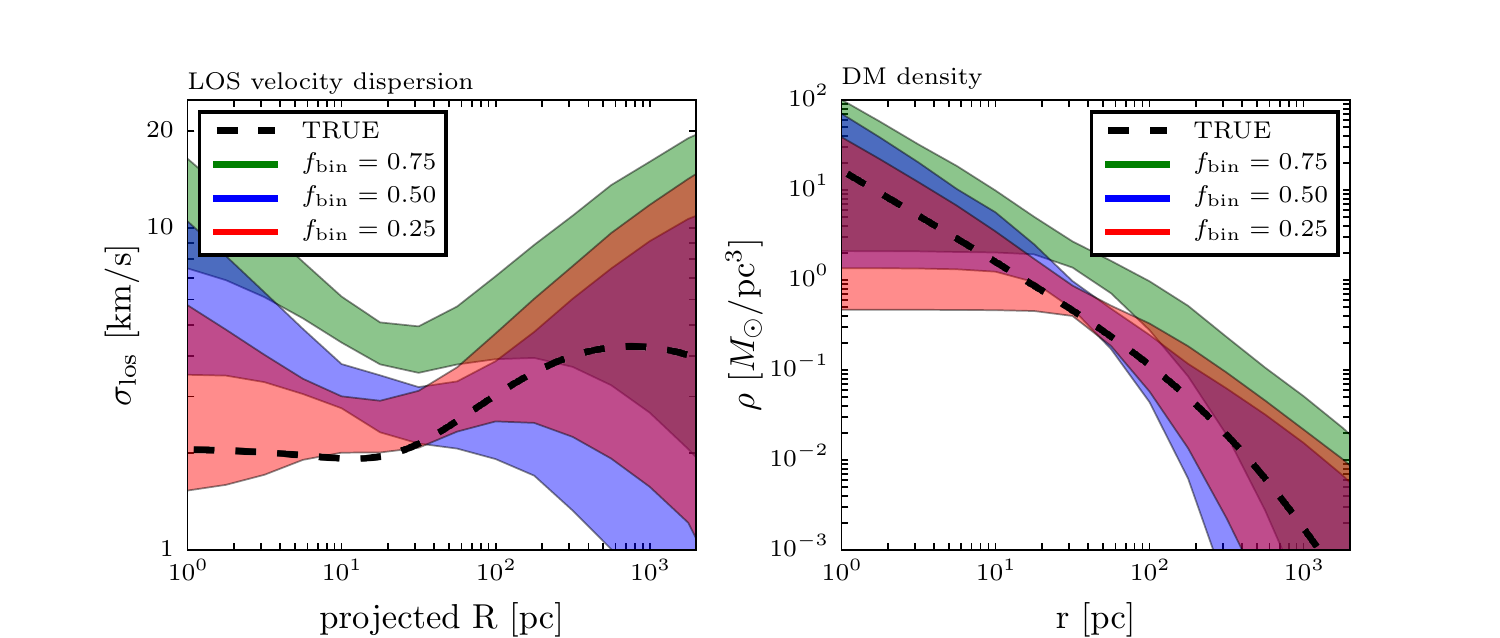}
\end{tabular}
\caption{Recovery of intrinsic line-of-sight velocity dispersion ({\it left}) and inferred dark matter density ({\it right}) profiles as a function of individual-star velocity precision ({\it top}) and the fraction of stars that belong to unresolved binary systems ({\it bottom}, with orbital parameters drawn from Galactic distributions), for an input model representative of low-mass ultra-faint satellites ($M_{200}=10^8M_{\odot}$, $R_{\rm half}=20$ pc).  
\label{f:uf_dsphmasses}}
\end{figure*} 

The relatively low spectral resolution at which most of the coldest ultra-faint systems have been observed compounds the sampling problems described above.  The multi-slit DEIMOS spectrograph at the Keck telescopes has resolution $R\sim 5000$, with a systematic error floor estimated at $\sim 2-3$ km s$^{-1}$ \citep{sg07}.  
Thus, the variability introduced by existing instrumental effects, as well as astrophysical effects like binarism, is similar to the intrinsic velocity dispersions measured for the coldest systems.  Using mock data sets, Figure \ref{f:uf_dsphmasses} shows the effects of both spectral resolution and unresolved binaries on measurements of velocity dispersion and dark matter density.  Clearly we require  velocity errors on single stars that are no larger than the intrinsic velocity dispersions to be resolved (top panels); the coldest known ultra-faint dwarfs exhibit dispersions near $\sim 2$ km s$^{-1}$ \citep{Caldwell2017}.  Furthermore, for such cold systems, failure to identify (and/or to model) binary stars generally leads to overestimation of velocity dispersions and dynamical masses.  Fortunately, given its unprecedented ability to observe faint targets across a wide field with high resolution at multiple epochs, MSE will not only provide samples of unprecedented size, but also unprecedented constraints on the above sources of systematic error.

\section{Galaxies in the low redshift Universe}\label{sec:lowz}

The CDM paradigm predicts the number, spatial distribution, and properties of galaxies over a wide range of masses.  While CDM predictions are in  remarkably good agreement with observations of high-mass galaxies ($M_\star > 10^{10} $M$_\odot$), studies of low-mass galaxies have raised several open questions \citep[e.g.,][]{Weinberg2013}. 
In particular, the low-mass end of the halo mass function and the profiles of low-mass dark matter halos are both especially sensitive to deviations from CDM. Current deep studies of dwarf galaxies either target the Local Group (a unique environment, see the previous section), or target relatively massive dwarf galaxies ($M_\star \gtrsim 10^{9}$\,M$_\odot$).  Spectroscopy is an essential ingredient to mapping the low-redshift ($z < 0.05$) dwarf galaxy population, linking galaxies to halos, and thus making inferences on the nature of dark matter.

New imaging surveys focused on the low-surface-brightness Universe are identifying large populations of dwarf and ultradiffuse galaxies outside the Local Group \citep{vanDokkum:2014cea,munoz2015,carlin2016,yagi2016,Geha2017,crnojevic2018,greco2018,smercina2018}, a discovery trend that will only accelerate with LSST.  However, their interpretation is severely hampered by a lack of distance measurements \citep{vanDokkum:2018vup,trujillo2018}.  Methods such as resolved stars and low surface brightness detections can identify dwarf galaxies out to $\sim 3$\,kpc and several tens of Mpc, respectively, with existing facilities \citep{Danieli2018}. Within $z < 0.05$,  dwarf galaxies have photometric properties that are very similar to background galaxies, and  photometric redshifts alone are uninformative in this regime (Figure \ref{fig:photoz}). 
Spectroscopic redshifts or tip-of-the-red-giant-branch distance estimates are critical for establishing distances for faint dwarf and ultradiffuse galaxies. At a minimum, a subset of spectroscopic-confirmed low-redshift galaxies are needed to calibrate photometric distance measures (e.g., using surface brightness fluctuations or training photometric redshifts).

\subsection{The faint end of the galaxy luminosity function}

A key prediction of $\Lambda$CDM is the hierarchy of halos down to small halo masses. Because galaxies are similarly hierarchical (large galaxies live in large halos, and small galaxies live in small halos), a comparison of the predicted dark matter halo distribution to the luminosity function of galaxies is an important test of $\Lambda$CDM.   Since there is some expected scatter in the galaxy -- halo connection, this test requires a large observed volume.  For brighter galaxies in the local Universe this has been possible with SDSS and GAMA \citep{driver2011}, however, it is below these masses where uncertainties arise. 

For galaxies fainter than $M_r \sim -14$, the uncertainty in the global low-redshift luminosity function is large, due to the incompleteness in both available photometric and spectroscopic surveys. This uncertainty in return results in large uncertainty in the galaxy--halo connection for halo masses below $10^{10} M_\odot$. Consequently, many inconsistencies between small-scale observations and model predictions \citep{Bullock2017araa} have not crystallized into concrete evidence for the need for non-cold dark matter due to the degeneracy between the galaxy -- halo connection models and dark matter models. 
In addition, a more complete sample of low-z galaxies also opens up the possibility of utilizing galaxy clustering statistics to further constrain galaxy -- halo connection models \citep{buckleypeter2018}. 

\begin{figure}[htb!]
  \centering\includegraphics[width=\textwidth]{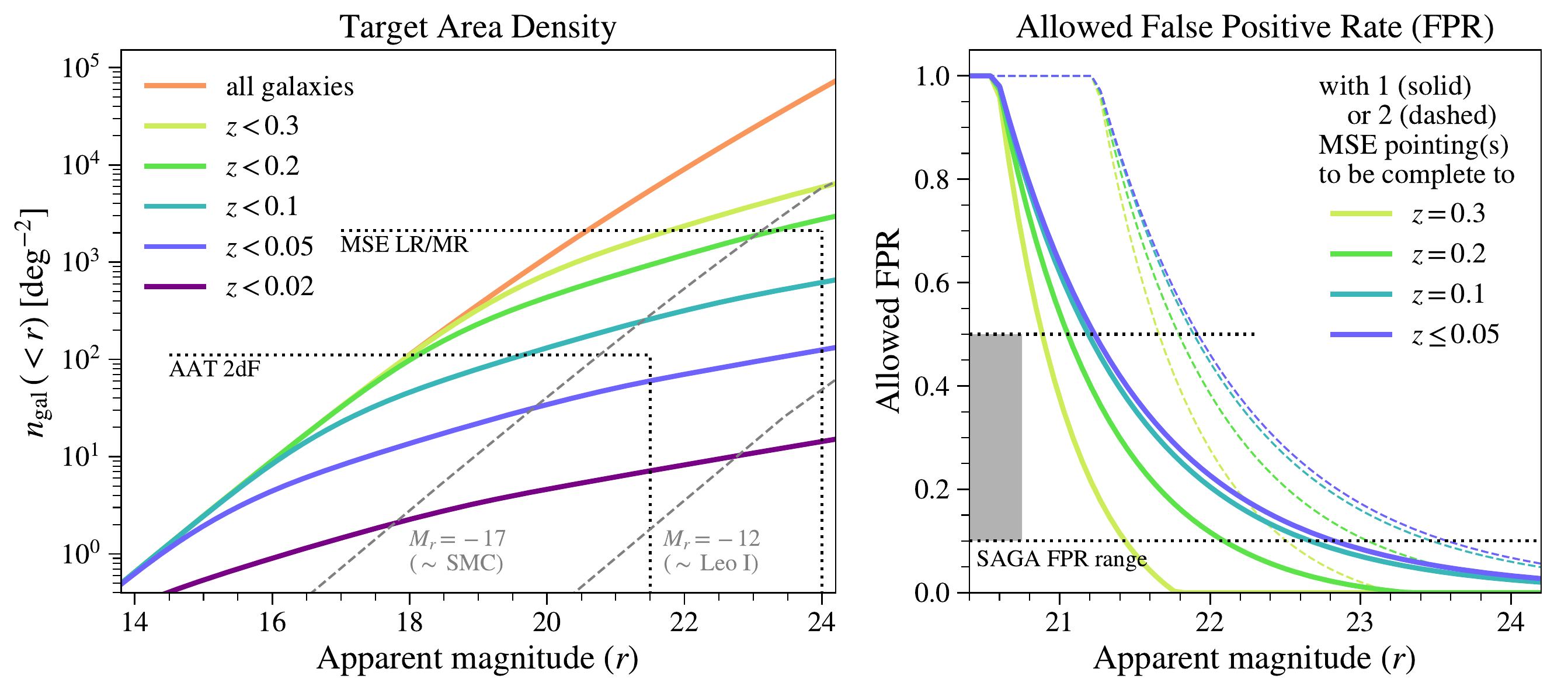}
  \caption{Left: Target area density (number of galaxies per squared degree; $y$-axis) as a function of apparent magnitude ($x$-axis) and redshifts (different solid color lines; redshift values shown in legend). The black dotted lines show the fiber area densities (horizontal) and magnitude limits (vertical) for AAT 2dF and MSE LR/HR.
  The grey dashed lines are isolines of equal absolute magnitudes ($-17$ and $-12$, roughly corresponding to the absolute magnitudes of SMC and Leo I, respectively).
  Right: Maximal allowed false positive rate (FPR) in target selection (the ratio of the number of targeted high-redshift galaxies to the number of fibers) to complete all galaxies below certain redshifts (different solid color lines; redshift values shown in legend) within only 1 (solid lines) or 2 (dashed lines) MSE pointing(s), as a function of the apparent magnitude of the galaxies. 
  The effect of fiber collision is not considered in this plot.
  FPR = 1 means that all high-redshift galaxies can be targeted and the low-redshift galaxy sample will still be complete; in other words, no extra photometric selection is needed in completing the low-redshift sample. FPR = 0.5 means that the photometric or morphological selection is needed so that only half of the fibers are targeted at high-redshift galaxies in order to complete the low-redshift sample. FPR = 0 means that the area density of low-redshift galaxies is equal to or higher than the fiber area density; therefore even if all fibers are used for the low-redshift galaxies, the sample is still not complete. 
  The horizontal black dotted lines (at FPR $=0.5,\,0.1$) are the range of SAGA FPR, i.e. in SAGA the high-redshift galaxy contamination is about 10\%-50\% after photometric or morphological selection (note that SAGA targets only go down to $r=20.75$).
  }
  \label{fig:lowz}
\end{figure}

Tightening the global low-redshift luminosity function is a critical, yet challenging, task.   In a given patch of sky, low-redshift galaxies are scarce due to the limited volume. To obtain a luminosity function that is as complete as possible, we need to measure redshifts of as many galaxies as possible. 
Figure \ref{fig:lowz} demonstrates the areal density as a function of redshift and apparent magnitude limits. In the left panel, we see that for $r < 20.5$, the low resolution fibers of MSE can obtain spectra for all galaxy targets in the field of view with just a single pointing (ignoring fiber collision). 
For $r < 24$, there are about 20 times more galaxy targets than the available fibers. In the right panel, we see that with an efficient target selection, MSE will be able to obtain a complete sample of low-z galaxies up to $z\sim0.1$ at $r < 24$ (c.f., the SAGA survey, \citealt{Geha2017}). 

\subsection{Satellite populations in Milky Way analogs}

Populations of satellite galaxies are particularly important probes of hierarchical formation models. The Milky Way is the most well-studied galaxy, and so far we have identified its $\sim 50$ dwarf galaxy satellites, including the Magellanic Clouds, the classical dwarf spheroidals, and more recently discovered ultra-faint dwarfs from SDSS and DES \citep{Bechtol2015,Drlica-Wagner:2015}. These satellites provide a unique probe of galaxy and star formation at early times in low mass objects, and also a powerful tool for distinguish different dark matter models.

However, the Milky Way satellite population constitutes a small, and perhaps biased, sample from which it is difficult to extrapolate.  For example, do host galaxies with similar luminosity, morphology, and mass as the Milky Way harbor a similar population of satellites?  Applying our detailed knowledge of the Milky Way satellites to broader questions of dark matter properties requires an improved understanding of satellite populations in the context of cosmology.

In fact, we know little about dwarf satellite galaxies outside the Milky Way and M31: the faintest detectable satellite galaxies around Milky Way analogs in SDSS (spectra to $r<17.7$) are similar to the Magellanic Clouds (MC). In SDSS, Milky Way analogs on average have only $\sim 0.3$ MC-like satellites,
v.s. two for the Milky Way \citep[e.g.,][]{Busha2011}. 
The recent SAGA Survey \citep[Satellites Around Galactic Analogs,][]{Geha2017} has taken on the exploration of finding satellite systems around Milky Way analogs. So far, the SAGA has constructed complete satellite luminosity function down to $M_r \sim 12$ (corresponding to Leo I), around 8 Milky Way analogs. This helps to constrain the intrinsic distribution of satellites around Milky Way mass galaxies, of which the Milky Way itself is a single realization.
However, a much larger sample is needed to reach any concrete conclusions. SAGA aims to obtain the satellite luminosity function for Milky Way mass galaxies between 20--40 Mpc, which amounts to about 100--200 hosts (depending on image availability). 
MSE will be able to push the boundary much further beyond 40 Mpc, or to push the satellite luminosity function to a fainter limit. As Figure~\ref{fig:lowz} demonstrates, MSE can easily obtain a satellite luminosity function down to $M_r \sim -12$ for any host galaxy within $z \sim 0.02$, provided a good target selection strategy, i.e. a false positive rate in target selection similar to SAGA.   

\begin{figure}[htb!]
  \centering\includegraphics[width=0.32\textwidth]{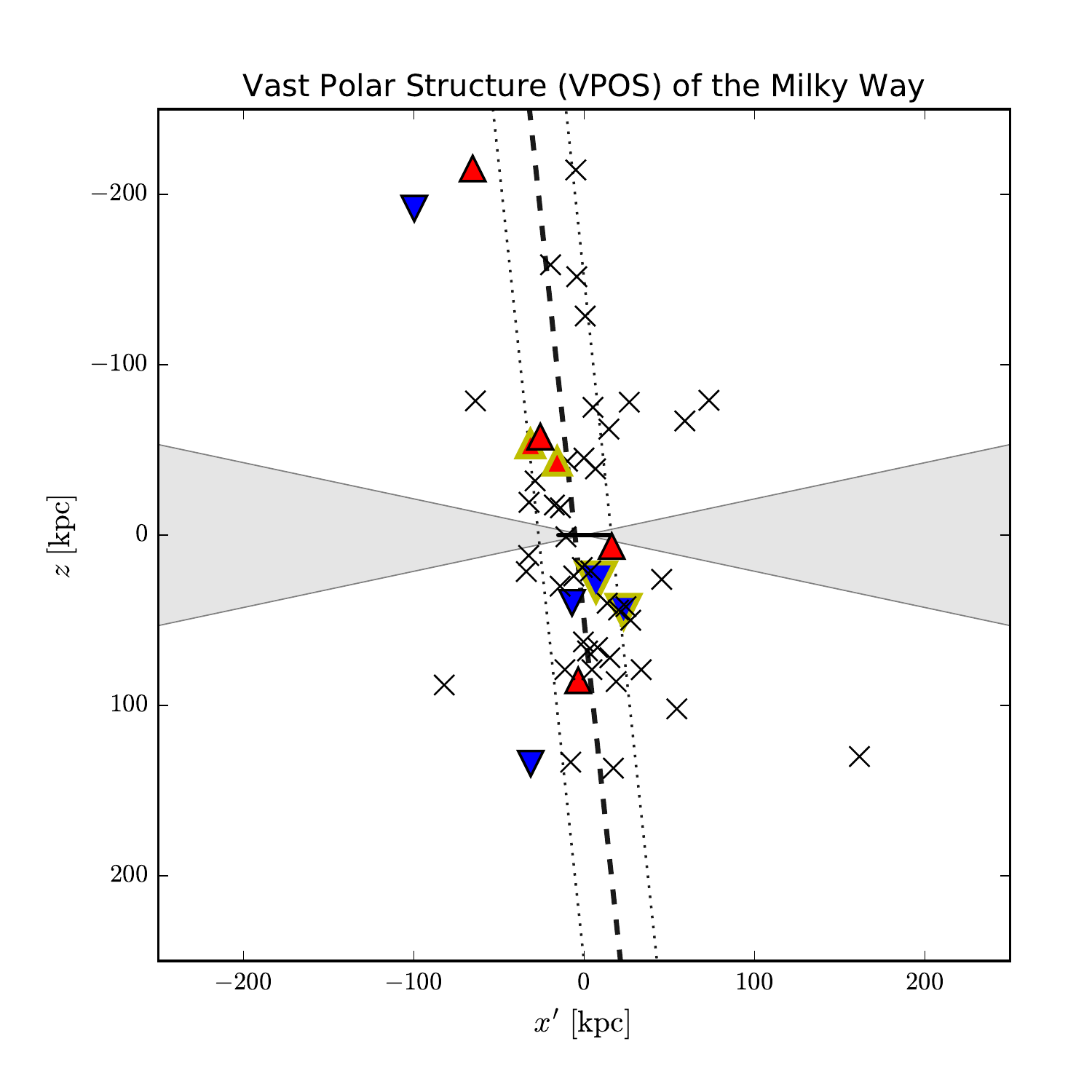}
\includegraphics[width=0.32\textwidth]{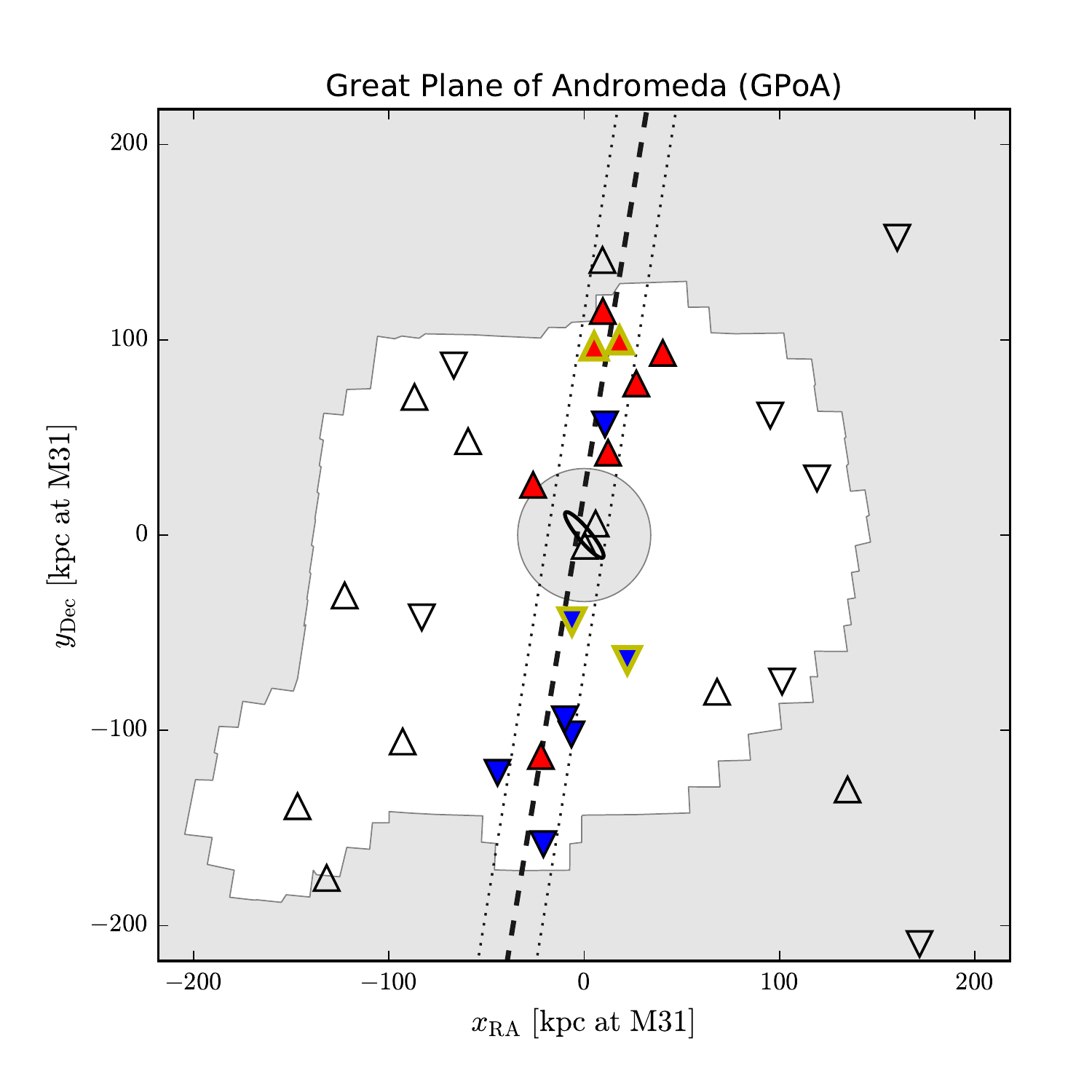}
\includegraphics[width=0.32\textwidth]{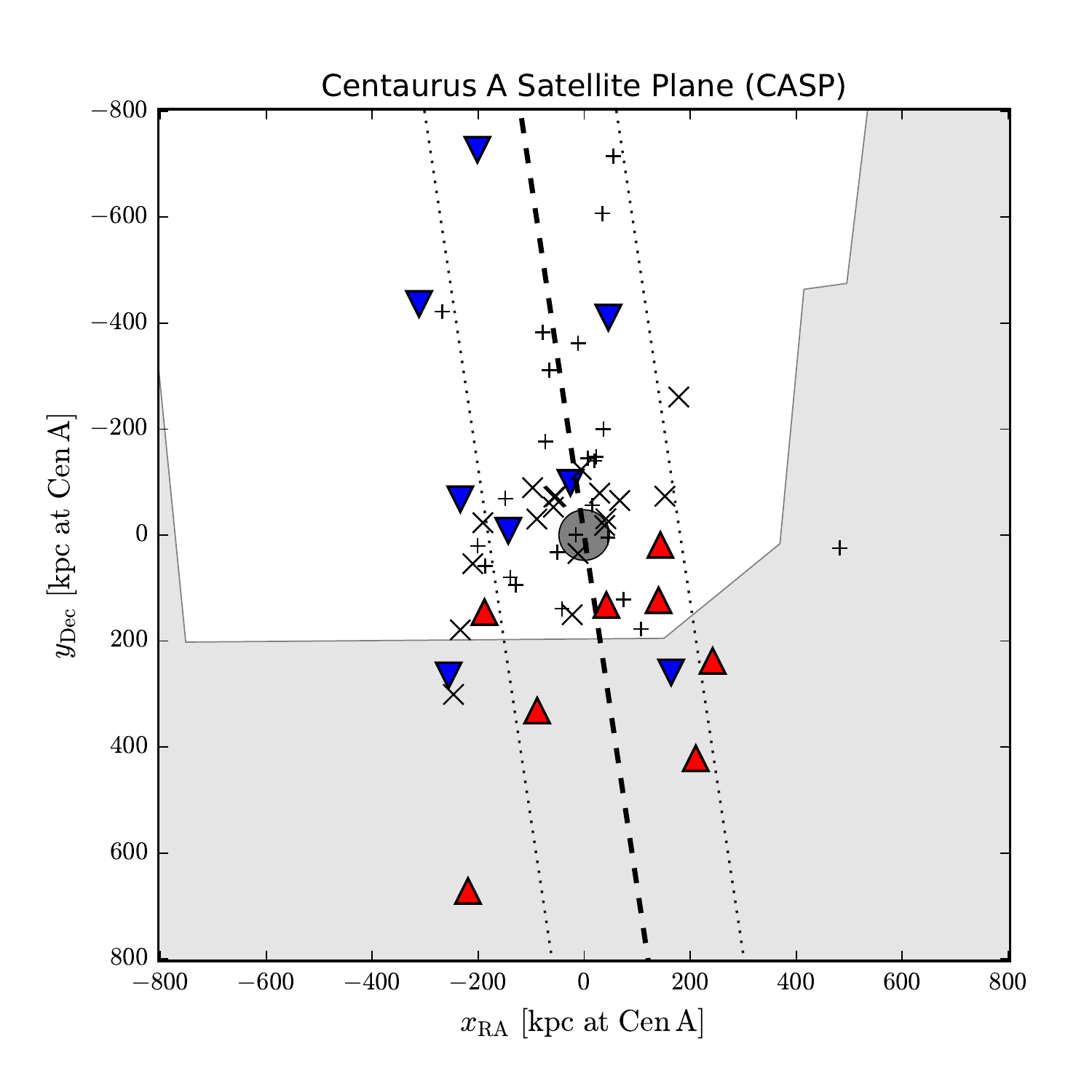}
  \caption{Edge-on views of three planes of satellites around the Milky Way ({\it left}), M31 ({\it middle}) and Centaurus A ({\it right}). The orientation of the flattened structures and their width are indicated with dashed and dotted lines, respectively. Satellites that are part of a planar structure and for which line-of-sight velocities are known (in the case of the 11 classical satellites of the Milky Way these are calculated from their proper motions) are shown as filled symbols, with upward red triangles indicating receding and downward blue triangles indicating approaching motion in these views. The two close pairs of satellites in the Milky Way and two in M31 identified by \citet{Fattahi2013} are highlighted with yellow outlines. Regions obscured by the Galactic disk, or outside of survey footprints, are indicated in grey. Figure from \cite{Pawlowski2018}.}
  \label{fig:planes}
\end{figure}

Comprehensive analyses of the satellite systems of Milky Way analogues should provide particularly important insights into recent results suggesting that planes of satellites exist around the Milky Way and other galaxies, possibly in conflict with predictions of CDM+simple galaxy formation models \citep[see Figure~\ref{fig:planes} and references in][]{Pawlowski2018}.  Indeed, testing whether such structures exist using a larger sample of satellites than currently exists offers a unique test of CDM. Satellite kinematics are less affected by baryonic effects: the positions and motions of satellite galaxies on scales of 100s of kpc is not strongly affected by their internal dynamics.   While this makes the issue particularly challenging to address within the $\Lambda$CDM framework, it holds potential to provide clues on the formation of dwarf galaxies and their accretion patterns that are not strongly dependent on the implementation of baryonic physics in simulations, but dominated by the overall dynamics governed by the dark matter distribution and properties.

Planes of Satellite Galaxies have two characteristic properties: (1) a spatial flattening in the positions of satellites around their host, and (2) a kinematic coherence, either seen in a preferred orbital direction (for the Milky Way where proper motions are available) or in line-of-sight velocities that are indicative of a rotating plane (i.e. satellites on one side are blueshifted, those on the other side are redshifted relative to the host). Obtaining spectroscopic systemic velocities for potential satellite galaxies around a larger sample of hosts is necessary to study both characteristics: 

\begin{enumerate}
\item The spatial analysis is improved by rejecting fore- and background contamination which can be assumed to contribute an isotropic signal that dilutes any spatial flattening. This is particular important since studying the spatial arrangement of satellite galaxies at distances beyond $\sim 5$\,Mpc loses one of three spatial dimensions; it is only possible in projection. This is because even small uncertainties of $\sim 5\%$, as achievable with the tip of the red-giant branch method, correspond to the whole virial volume of a Milky Way-like host;

\item While full 3D velocities would be required to confirm or refute a rotational support of found satellite planes, line-of-sight velocities alone can already give a statistical answer to the prevalence of kinematically coherent structures. For random sight-lines, a plane of satellite galaxies is seen to higher than $60^\circ$ inclination in 50\% of the cases, which implies that the line-of-sight velocity is dominated by the in-plane component if the satellite plane is indeed rotating.
\end{enumerate}

More generally, a larger sample of satellite systems with spectroscopically confirmed satellite galaxies allows studies of other, potentially related, phase-space correlations among satellites. One example is the apparent over-abundance of pairs of satellites in the Local Group compared to $\Lambda$CDM simulation \citep{Fattahi2013}, another is the abundance and properties of groups and associations of dwarf galaxies pre- and post-infall onto a host galaxy (e.g., \citealt{sales2011, sales2017} and references therein).   A third example is the alignment of satellite planes with their surrounding large scale structure, which could give clues on their origin. Current studies \citep{Libeskind2015} give mixed results -- the planes around M31 and Centaurus A appear to align with the ambient shear field, that around the Milky Way does not -- but these studies are severely limited by the small sample size of known satellite planes.   MSE can provide a much larger sample to investigate these questions in a statistically meaningful way.

\subsection{Local galaxies as gravitational lenses}

Above halo masses of $10^{11} M_\odot$ (corresponding roughly to an LMC-mass galaxy), there is good agreement among various probes of the relationship between the stellar mass and halo mass  \citep[e.g.,][]{Wechsler:2018pic}.  Below this mass scale, measurements of the relation based on dynamical masses within the optical radius of galaxies (e.g., with rotation curves) and based on abundance matching diverge, with dynamical measurements preferring smaller halo masses \citep[e.g.,][]{klypin2015,papastergis2015,buckleypeter2018}.  This mismatched is sometimes called the ``Too Big to Fail in the Field" problem.  The fundamental problem is that neither of these measurements of halo mass is direct, but instead rely on extrapolations.  In the case of dynamical measurements, the match between the stellar mass and halo mass depends on strong priors about the density profile to connect the optical radius to the virial radius (a factor of $\sim 10 - 100$ in scale).  In the case of abundance matching, the key assumptions are that the halo mass function is the CDM halo mass function, and that all halos at these masses should have galaxies inside of them.  The field needs a direct measurement of the halo mass of objects of Magellanic Cloud-mass or below in order to tell if the mismatch in the indirect measurements is a result of a density profile problem, a halo count problem, or a star-formation stochasticity problem. To this end, weak (gravitational) lensing provides a direct unbiased measurement of the total mass, and of the density profile of the halo outside the optical radius.   

Weak lensing refers to the subtle distortion of galaxy shapes due to light being deflected from the large-scale matter distribution in the line-of-sight \citep{Bartelmann2001}. The weak lensing signal is typically detected statistically as the average distortion of the galaxies are very weak (sub-percent compared to average intrinsic galaxy shapes at about 30 percent). Galaxy-galaxy lensing refers to a specific statistics used to extract this signal, specifically a correlation of the foreground (lens) galaxy position and the background (source) galaxy shape distortion, or shear. This effectively gives a measure of the average mass distribution around the foreground galaxy sample. With sufficient signal-to-noise, one can use weak lensing to constrain the total mass of the low-mass galaxies described above. \citet{Sifon2018} showed a proof of concept using 784 ultra-diffuse galaxies (UDGs, see more discussion in Section~\ref{sec:udgs}) around 18 clusters at $z < 0.09$ and obtained an upper bound for the average mass of the UDGs. Similar measurements can be done with dwarf galaxies or other specific low-mass lens samples.

One further possibility is to use the full weak lensing profile shape (in addition to the amplitude) to constrain different dark matter models. Baryonic and dark matter physics can both alter density profiles from their NFW-like baryon-free CDM form \citep{Dave:2000ar,Colin:2000dn,Governato:2009bg,Fitts:2018ycl}. These effects may be separable (although baryons lead to ``convergent evolution" of halo properties relative to their baryon-free predictions), and so measuring deviation of the profiles from NFW could provide constraints on e.g. SIDM models \citep{Ren:2018jpt}.  Moreover, a measurement of the density profile can help determine the physical origin of the discrepancy between dynamics and abundance-matching-based inferences of the stellar-mass--halo-mass relation.

\begin{figure}
  \centering
    \includegraphics[width=\textwidth]{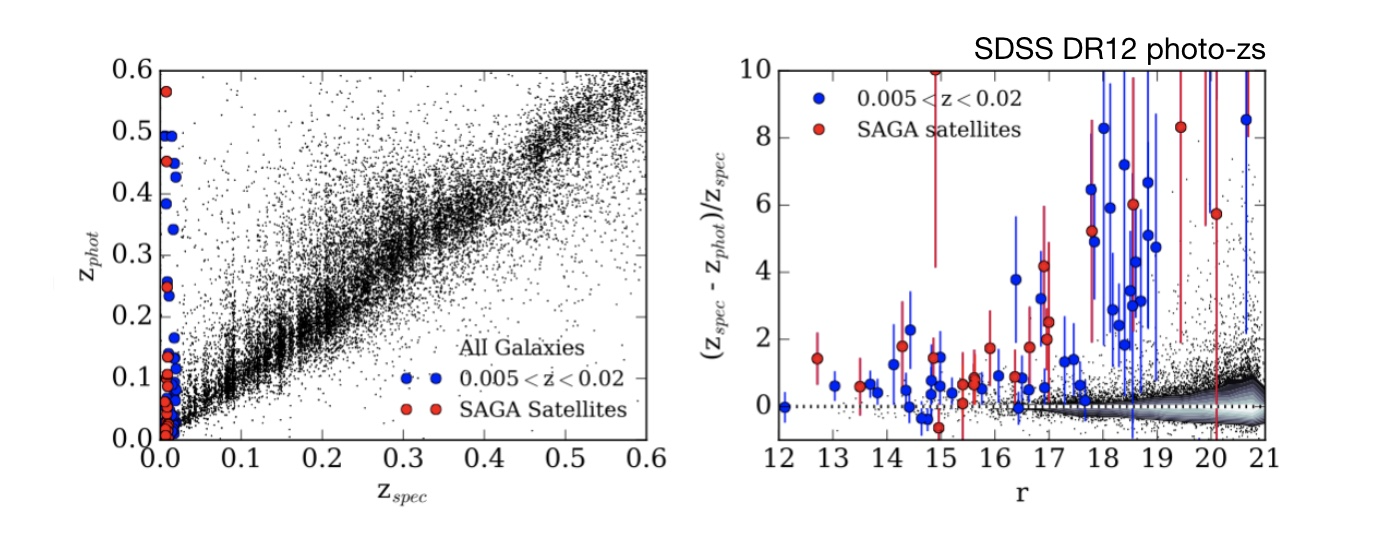}
      \caption{({\it Left}\/) Spectroscopic redshift plotted against the SDSS DR12 photometric redshifts from \citet{Beck2016}. ({\it Right}\/)  Apparent $r$-band magnitude versus the fractional difference between the spetroscopic and photometric redshifts.  In both panels, we plot SAGA satellite galaxies as red circles and a larger number of field galaxies over a similar redshift range ($0.005 < z < 0.015$) as blue squares.   For the majority of galaxies with redshift $z < 0.015$, particularly for galaxies fainter than $r_o > 17.7$, photometric redshifts are neither accurate nor precise. Figures from \citet{Geha2017}.}

  \label{fig:photoz}
\end{figure}

The two main challenges of these measurements are (1) the number and average redshift of these galaxies are low, and so we expect the signal to be relatively weak; (2) contamination of high-redshift galaxies in the lens sample could introduce spurious signal and result in a bias in the weak lensing mass/profile inferred from the data. Having accurate spectroscopic redshifts from MSE will help for both of these aspects compared to the case where only photometric redshift (photo-z) estimates are available. First, even assuming optimistically that the photo-z estimates are unbiased, the scatter would smear out the signal and lower the detection significance. Second, having spectroscopic redshift ensures a much cleaner lens sample without catastrophic photo-z outliers that could bias the inferred weak lensing mass. Figure~\ref{fig:photoz} demonstrates the potential problems of using photo-z estimates for the weak lensing measurements of the low-redshift galaxies. At low redshift ($z<0.015$) and faint magnitudes ($r>17.7$), where the population of interest lies, one can see a significant fraction of outliers in the photo-z estimates and large scatter.

\subsection{Ultra diffuse galaxies}\label{sec:udgs}

The recent research focus on very low surface-brightness galaxies (so-called “Ultra-Diffuse Galaxies” or UDGs) in nearby galaxy clusters has opened a surprising new avenue to investigate the galaxy-halo connection. While such objects have been known to exist for decades \citep{sandage1984,dalcanton1997}, their abundance in clusters was not fully appreciated; hundreds are now known to live in the most massive clusters \citep{koda2015,vanderburg2017}.
Notably, the few UDGs for which spectroscopic studies have been conducted exhibit unusual properties associated with their dark-matter halos, such as large globular cluster populations \citep{vandokkum2016} or very high \citep{vandokkum2016,beasley2016} or low velocity dispersions \citep{vandokkum2018}. Although many explanations have been put forth to explain these unusual galaxies, none are able to completely reproduce the observed properties \citep[e.g.~][]{amorisco2016,chan2018,carleton2018}. Further spectroscopic observations are necessary to both directly probe their dark-matter halos and develop a more complete understanding of their formation and evolution. As these unusual galaxies primarily live in group and cluster environments \citep{vanderburg2017}, they are natural targets for the high multiplex of MSE.

Low-surface-brightness galaxies have long been understood as excellent laboratories for understanding the galaxy -- halo connection, as their dynamics are governed by their dark matter halos both at small and large radii. However, current studies arrive at vastly divergent conclusions regarding the dark-matter content of UDGs, with reported halo masses spanning 4 orders of magnitude \citep{vandokkum2016,Sifon2018,vandokkum2018}. While indirect measurements suggest that UDGs typically live in dwarf halos \citep[e.g.~][]{Amorisco2018}, direct spectroscopic measurements persistently find unusually high \citep{vandokkum2016,beasley2016} or low \citep{vandokkum2018} halo masses. In the latter case, follow-up observations paint an even more complex picture; globular clusters and planetary nebulae are both consistent with a $10$\,km\,s$^{-1}$ dispersion \citep{Laporte2018,Martin2018,Emsellem2018}, whereas the more concentrated stellar component has a slightly higher dispersion at $16$\,km\,s$^{-1}$ \citep{Emsellem2018}.
A larger sample of UDGs with direct mass estimates from a variety of tracers is crucial to resolving this apparent discrepancy as only four galaxies have been spectroscopically followed-up (in terms of their globular cluster abundance or stellar velocity dispersion). Regardless, these unexpected results indicate that further study of the dark-matter content of UDGs will be particularly fruitful for our understanding of the physics of dark matter and the role it plays in galaxy evolution.

These measurements will be possible with the large field-of-view and multiplexing capabilities of MSE. While direct stellar velocity dispersion measurements will only be possible for a few UDGs with dozens of hours of observations, MSE will measure the velocity dispersion of the associated globular cluster population for all UDGs out to the distance of Virgo. Additionally, spectroscopic observations of UDGs in clusters as far out as Coma with MSE will provide spectroscopic distance measurements necessary to confirm the large sizes and cluster membership of UDGs, paving the way for deeper follow up studies. Current samples of UDGs suffer from significant foreground contamination \citep{vanderburg2016}, and simple spectroscopic redshifts obtained with MSE will greatly improve the sample of UDGs that are spectroscopically confirmed. For example, this spectroscopic identification will greatly improve weak lensing constraints on UDG masses \citep{Sifon2018}, in a similar way that it will improve weak lensing constraints on dwarf galaxies in general (see previous section). Furthermore, stellar metallicities and ages of both the UDGs and their globular clusters will provide valuable checks on theories for UDG formation and evolution \citep[e.g.~][]{Gu2018}. In particular, robust identification and measurement of the rich UDG globular cluster population will allow for a better understanding of this least-understood aspect of UDGs.

The capabilities of MSE place it in an excellent position to address these questions regarding UDGs. MSE’s large field of view allow it to cover the $\sim 300$ UDGs in Coma in less than 10 pointings. While the central surface brightness of UDGs, ranging from $23 - 25$ mag/arcsec$^2$, push the limit of MSE, a spectroscopic census of UDGs down to $25$ mag/arcsec$^2$ is possible. Deeper observations will provide high-quality spectra necessary to measure metallicities and ages of the stellar populations of UDGs, as well as direct stellar dispersions for a sample of brighter UDGs. Given the wealth of data obtained through spectroscopic observations of individual UDGs, this type of survey will transform our understanding of the low surface brightness Universe. Additionally, MSE is a particularly powerful instrument to target globular clusters around UDGs in Virgo, as it will be able to identify and the velocity dispersion of $10 - 20$ globular clusters (with $r$-band magnitudes spanning $20 - 24$) around the very extended (half light radii $>1$\,arcmin) UDGs in Virgo in a single pointing. Spectra of these objects with velocity resolution better than several km\,s$^{-1}$ will provide a robust constraint on the dark-matter content of these galaxies, provided enough tracers are present \citep[][]{Laporte2018} or at least spectroscopically confirm globular cluster candidates. 

\section{Galaxies beyond the low redshift Universe}\label{sec:highz}

MSE can measure the dark matter distribution and the halo mass function in galaxies beyond the low-redshift ($z>0.05$) Universe. We consider three different probes in the following sections that show promise for constraining the particle nature of dark matter. Two of these are based on gravitational lensing effects of small halos and they will both measure the halo mass function to scales smaller than $10^8 \, M_\odot$. The third is based on kinematics of the bright cluster galaxies close to the center, with the power to constrain the smooth halo profiles of clusters of galaxies. 

\begin{figure*}
\centering
\scalebox{0.45}{\includegraphics{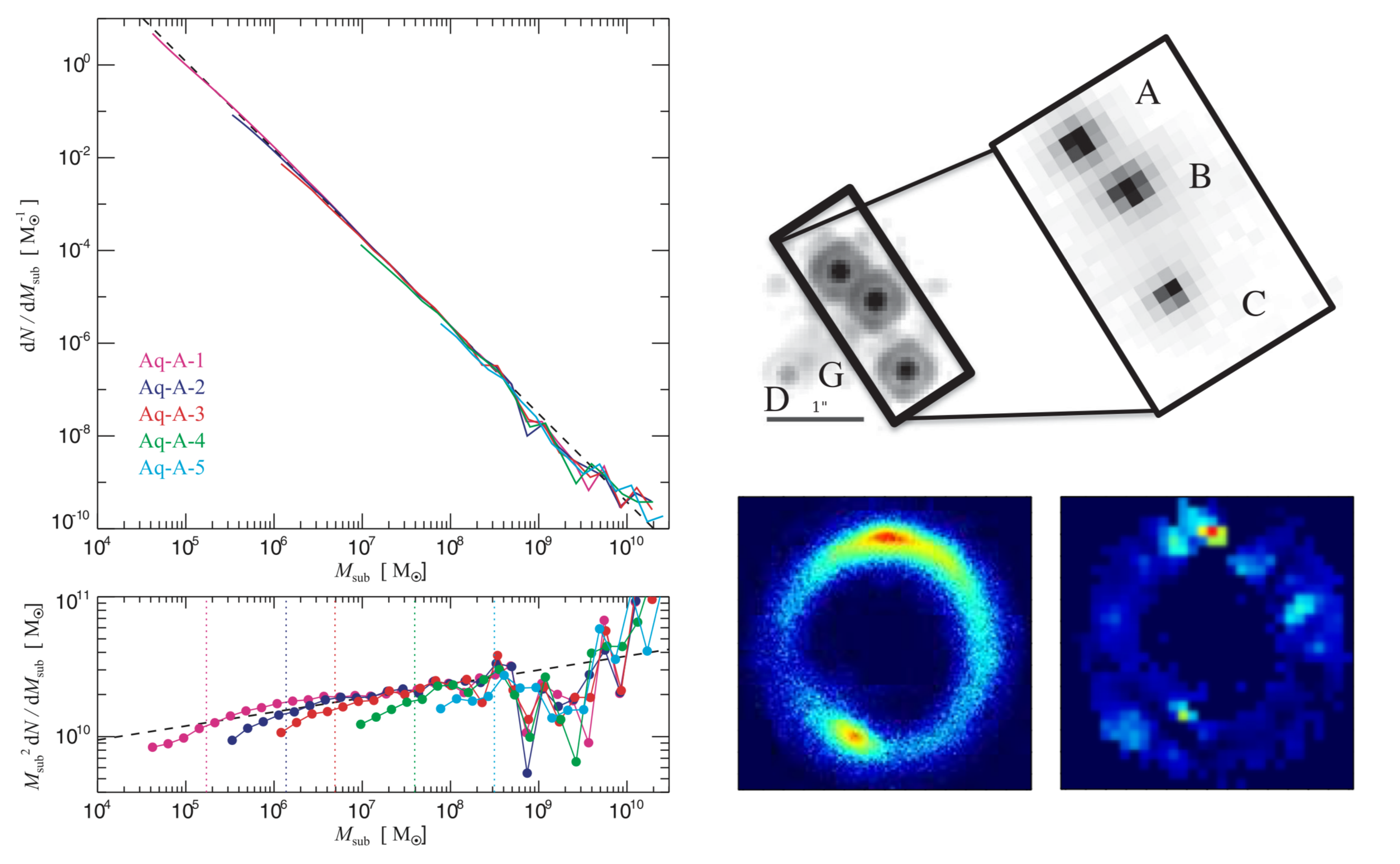}}
\caption{Left: subhalo abundance within simulated Milky Way-scale halos from \citet{springel_etal_2008}. Integrated abundances are of the order 1,000 subhalos in the decade between 10$^7$ and 10$^8$\,$M_{\odot}$, and of the order 100 subhalos in the decade between 10$^8$ and 10$^9$\,$M_{\odot}$. Gravitational lensing is sensitive to perturbations from these subhalos as well as all halos along the line-of-sight \citep[e.g][]{Despali++19, Gilman++18}. Upper right: HST-NICMOS imaging and Keck-OSIRIS integral-field data for the quadruply lensed quasar B1422$+$231 from \citet{Nierenberg++14}, used to detect the presence of substructure in the lens galaxy through flux-ratio anomalies. Lower right: Keck NIRC2 adaptive optics-assisted image of the lensed galaxy B1938+666 from \citet{Vegetti2012}, along with reconstructed density perturbation from a dark-matter dominated satellite to the lens galaxy detected via surface-brightness perturbations to the Einstein ring image. 
\label{f:L-star-lensing}}
\end{figure*}

One of the most robust predictions of the $\Lambda$CDM model is the ubiquity of hierarchical mass substructure at all scales down to the free-streaming cutoff length of the dark matter particle \citep[e.g.][Figure~\ref{f:L-star-lensing}, left]{springel_etal_2008}. On the scale of individual massive galaxies, the absence of dwarf satellite galaxies in comparable abundance could be considered a challenge to the $\Lambda$CDM paradigm, or alternatively could just be a reflection of mass-dependent star-formation efficiency and observational selection processes \citep[e.g.,][]{klypin_etal_1999}. Strong gravitational lensing on the scale of individual galaxies provides a  method to constrain substructure directly in the dark sector and beyond the local Universe, since lensing is sensitive to all gravitating mass independent of luminosity. For unresolved sources such as lensed quasars, this method operates through the detection of flux-ratio anomalies: differences between the relative magnifications of lensed images as compared to the predictions of smooth mass models \citep[e.g.,][]{Dalal++02}. For resolved sources such as lensed normal galaxies, the method operates through the detection of surface brightness perturbations \citep[e.g.,][]{Vegetti2010} associated with the presence of gravitating substructure in the lens galaxy. The right panel of Figure~\ref{f:L-star-lensing} illustrates both these cases.

The large number of new galaxy-scale strong lenses that will be delivered by MSE (both alone and in conjunction with imaging facilities) will enable strong-lensing tests of dark-matter substructure at high significance, providing a fundamental test of the $\Lambda$CDM hypothesis through constraints on the parameters of the dark-matter halo and subhalo mass functions. We consider the prospects for these experiments through the flux-ratio anomaly and surface-brightness perturbation channels in the following sections. We then discuss a novel technique to infer the dark matter distribution in the inner regions of clusters via the wobbling of the brightest central galaxies (BCGs). 

\subsection{Quasar lensing: flux ratio anomalies due to low mass dark matter halos}

In strong gravitational lensing, multiple images of a background source appear due to the distortions in space-time caused by one or more intervening massive objects along the line of sight to the observer. The positions and relative brightnesses of these multiple images depend on the
first and second derivatives of the gravitational potential of the deflector, respectively. Due to this dependency, the image positions provide a strong constraint on the smooth, larger-scale mass distribution of the deflector, while the relative image magnifications are extremely sensitive to low-mass halos. The current limit is $M_{200}\sim 10^{6.5} $M$_\odot$ with current technology \citep[e.g.][]{Nierenberg++14, Nierenberg++17}.

This method was first applied to strong lenses two decades ago \citep{Mao++98}, but it has not yet reached its full potential, because the number of suitable quasar lenses has been too small. In particular, useful quasar lenses must have {\it four} images to provide an accurate constraint on the smooth mass distribution. Additionally, the source must be at least milli-arcseconds in size to avoid significant perturbations by stars in the plane of the deflector. These two requirements have heretofore limited the field to the small number of currently known radio-loud quasar lenses \citep{Dalal++02}. New technologies have recently made it possible to extend this analysis to more systems by measuring strongly lensed quasar narrow-line emission, which is observed in virtually all quasars (unlike radio). This is extremely promising: nearly all optically selected quasars have significant narrow-line emission, making it possible to extend the strong-lensing measurement of the dark matter mass function to many more systems  \citep{Moustakas++03}. 

Based on recent, state-of-the-art simulations \citet{Gilman++18}, it is estimated that, with approximately 100 lenses (depending on flux precision), it will be possible to place a more stringent constraint on the `warmth' of dark matter --- compared to the constraints provided by the Ly-$\alpha$ forest \citep{Viel++13}. With several hundreds of lenses it will be possible to rule out even lower-mass cutoffs in the power spectrum. Such a measurement will have completely independent systematic uncertainties and therefore provide a crucial probe of dark matter below the mass scale at which dark matter halos are currently known to reliably form galaxies.
 
LSST will contain $\sim1000$ four-image quasar lenses, in which three images will be brighter than the survey magnitude limit \citep{Oguri++10}. With current gravitational lensing techniques and IFU sensitivity alone, this number will be sufficient to provide stringent new constraints on a turnover in the dark matter power spectrum \citep{Gilman++18}. MSE will play two crucial roles in this constraint both by confirming many of the quasar lenses, and by selecting ideal candidates for follow-up with the next generation of 30-meter class telescopes. 
MSE will provide an essential step in reaching this goal of measuring microlensing-free fluxes for hundreds of quasar lenses.     

\begin{figure*}
\centering
\includegraphics[width=2.5in,trim=0in 0in 0in 0in,clip]{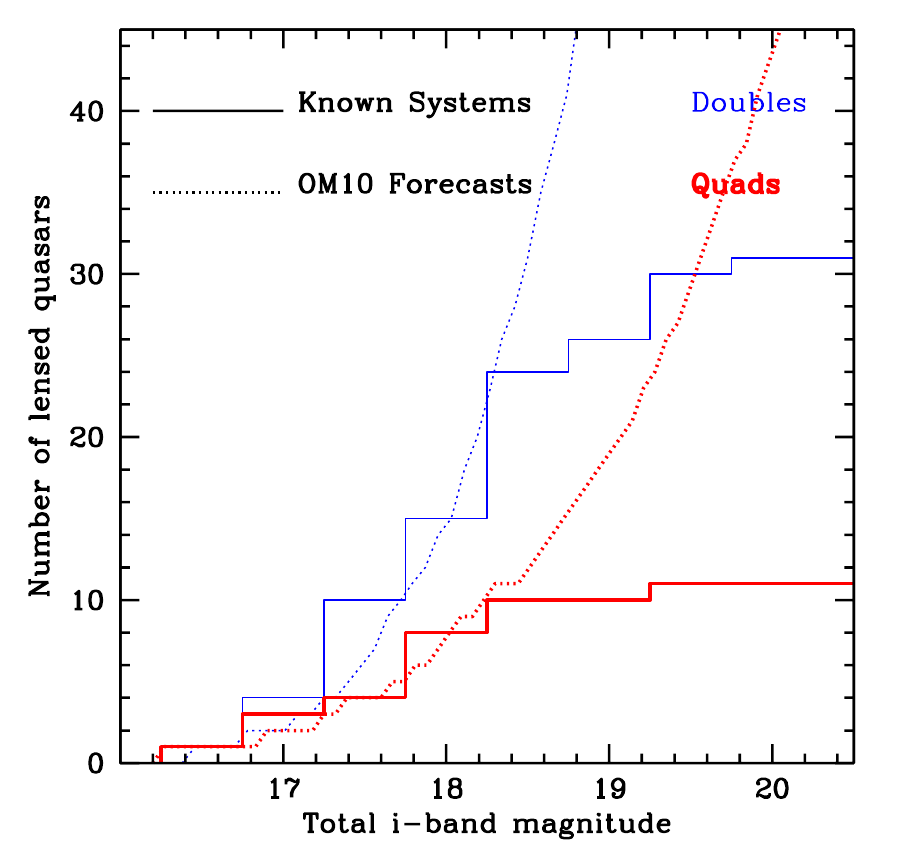}
\includegraphics[width=3.5in,trim=0in 0in 0in 0.in,clip]{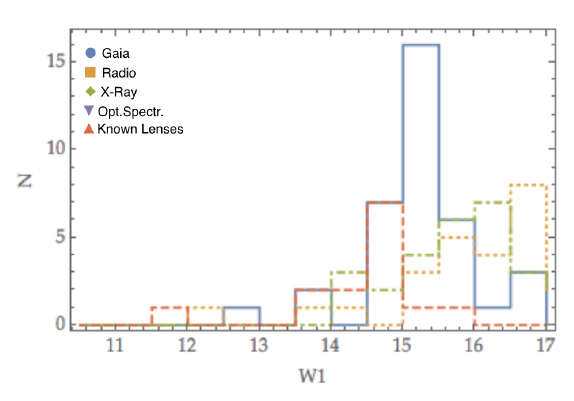}
\caption{Detection of quasar lenses in optical data relies heavily on ancillary data. In the case of DES it relies particularly on infrared data from WISE. Left: The number of quasar lenses from \citet{Oguri++10} predicted to be found in the DES footprint (dotted lines) compared with the observed numbers (solid histogram). Lens finding algorithms become incomplete well below the depth of the DES survey ($\sim 24$ i-band). Figure from \cite{Treu++18}. Right: Distribution of WISE W1 magnitudes of quasar lenses and lens candidates. The candidate numbers peak at the limiting magnitude WISE W1 filter. In order for optical surveys to reach their full potential for gravitational lensing, deep ancillary data is crucial. Figure from \cite{Agnello:2017}.
\label{f:DEScompleteness}}
\end{figure*}

Deep, high-resolution imaging will enable morphological and color selection of quasar lens candidates. However, based on results from \citet[e.g.][]{Agnello++15} and \citet[e.g.][]{, Agnello++18}, color and morphological information alone is insufficient to separate quasar lenses from the ``blue cloud'' of galaxies. For example, lens searches in DES have relied on WISE infrared photometry to isolate objects with quasar-like colors. As is shown in Figure \ref{f:DEScompleteness}, the number of identified four-image lenses is incomplete relative to theoretical predictions at magnitudes fainter than an \emph{i}-band magnitude of 18,  which is far below the survey depth of $i \sim 24$. This is due to the limit imposed by requiring WISE photometry, as shown in the right hand panel of Figure \ref{f:DEScompleteness}. In contrast, in SDSS, matching spectroscopy enabled the discovery of quasar lenses down to the limiting survey magnitude of $i \sim 21$ \citep{Inada++12}. MSE can provide critical spectroscopy for this science in conjunction with current and future imaging surveys.

For the goal of measuring a turnover in the low mass end of the halo mass function, we conservatively require approximately 200 lenses \citep{Gilman++18}. Over the area of LSST, this can be achieved by reaching a limiting spectroscopic depth of $i \sim 22$ \citep{Oguri++10}.  The number of quasar candidates can be estimated with a purely optical color selection using the results from \citet{Richards++2002}, which identified approximately 18 candidates per square degree to a limiting $i$-band magnitude of 19 with SDSS photometry. Given a true number at this magnitude of approximately one quasar per square degree, and making the coarse assumption that the purity remains constant with magnitude, we therefore expect about 600 quasar candidates based on optical color selection for the 30 true quasars with $i < 22$ per MSE pointing. This is certainly feasible given the $> 3000$ low resolution fibers for MSE. However, the number can likely be further reduced significantly with the addition of morphological cuts.

In addition to confirming quasar lenses and to measuring their source redshifts, MSE will enable the measurement of spatially-blended, narrow-line fluxes. Given the fiber size, spectra from all four images of a lensed quasar will be blended into a single measured spectrum. This narrow-line flux measurement will enable accurate planning for follow-up observations with higher spatial resolution facilities such as TMT, and enable the elimination of targets with significant spectral contamination from broad iron or hydrogen lines. 

\subsection{Galaxy-galaxy lensing: image perturbations by low mass dark matter halos}

\begin{figure*}
\centering
\scalebox{0.3}{\includegraphics{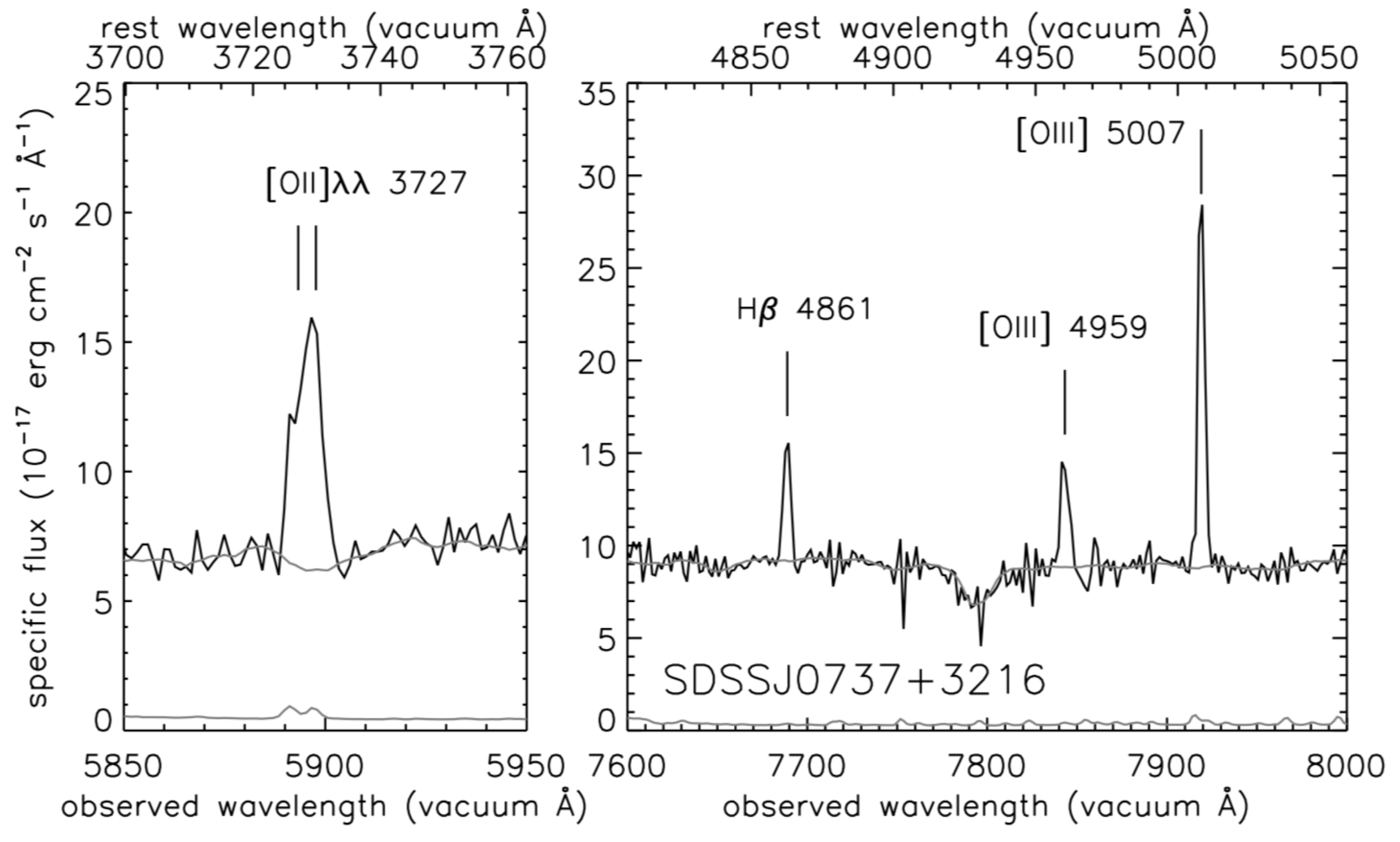}}\scalebox{0.35}{\includegraphics{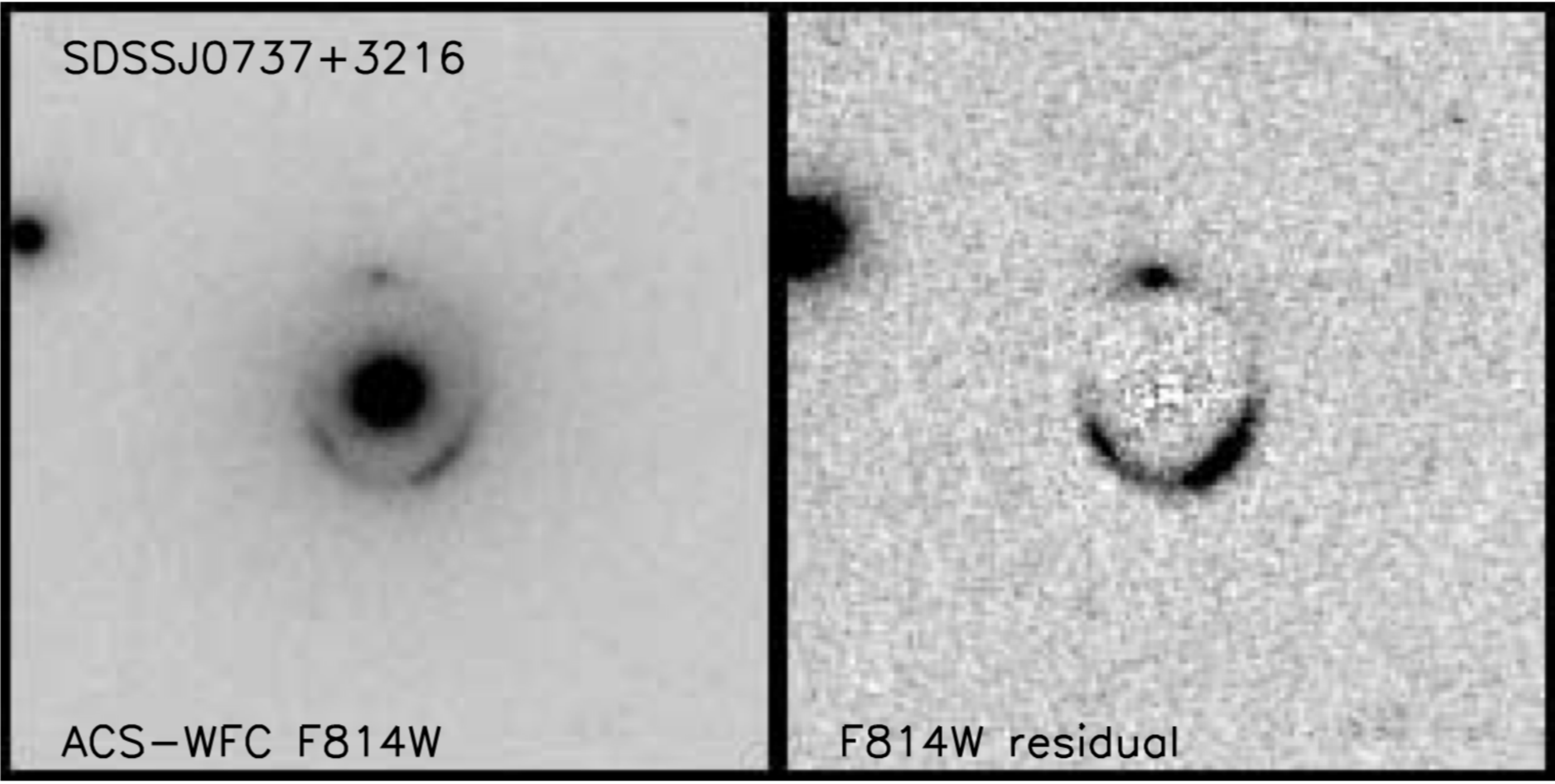}}
\caption{Spectroscopic selection of strong galaxy-galaxy gravitational lenses in the Sloan Lens ACS (SLACS) survey. Left: [O\textsc{ii}], H$\beta$, and [O\textsc{iii}] emission lines superimposed on the spectrum of a lower-redshift quiescent galaxy. Right: HST imaging of this same system, confirming it to be a gravitational lens. Figure from \citet{Bolton2006}.
\label{f:SLACS}}
\end{figure*}

Galaxy redshift survey spectra have proven to be an unparalleled resource for the discovery of new strong galaxy-galaxy lensing systems in large numbers. Through analysis of multiple generations of the SDSS, it has been found that $\sim0.1 - 1\%$ of galaxy spectra show evidence for another galaxy (or quasar) along the same line of sight. Follow-up imaging reveals a substantial fraction of these `lens candidates' to be bona fide strong gravitational lenses (e.g., \citealt{Bolton2006, Bolton2008, Brownstein2012, Shu2017, Shu2018}). This technique is illustrated in Figure~\ref{f:SLACS}.

The sensitivity and multiplexing capability of MSE, combined with its dedicated survey operations mission, can enable flux-limited galaxy surveys ten times larger than the original SDSS.
These surveys can be expected to deliver significant new samples of strong galaxy-galaxy lenses, which can in turn enable high precision lensing tests of the low mass end of the CDM halo mass function.

In a strong galaxy-galaxy lens system, dark matter perturbations to the smooth mass distribution of the lensing galaxy can be detected via astrometric perturbations to lensed background sources, which appear as distortions in lensed arcs \citep{Vegetti++09, Vegetti++2010a, Vegetti++2010b, Vegetti++12, Hezaveh++16}.
Both the subhalos of the main lens halo and the halos along the line-of-sight lead to these distortions, with the latter dominating at higher redshifts \citep{Despali++19}.
The mass sensitivity of this method depends on the depth and spatial resolution of the available imaging data, an accurate model of the intrinsic unperturbed source light distribution, and the redshifts of the lens and source galaxies. 
Currently, the method is not as sensitive to the small-scale power spectrum \citep{Vegetti++18, Ritondale++19} as Ly-$\alpha$ forest measurements \citet{Viel++13} but it shows great promise for future applications given the expected sensitivity to dark matter halos of virial masses as low as $\sim 10^8 M_\odot$ \citep{Vegetti2014}.

In order to improve the constraints on the small-scale dark matter power spectrum with `gravitational imaging' of lensed galaxies at optical and IR wavelengths, a combination of improved imaging precision, larger sample sizes, and selection of a set of `ideal' galaxy-galaxy lenses is required. While the first requirement can only be met by deep, high-resolution imaging facilities such as HST, JWST, and AO-assisted instruments on large-aperture ground-based telescopes, the latter two requirements are directly addressed by the aperture and spectroscopic multiplex capacity of MSE. The third requirement is especially relevant in consideration of the depth that MSE will achieve relative to SDSS, since higher redshift sources probe a longer path through the Universe and thus provide a tighter constraint on CDM \citep{Despali++19}. The discovery of many new galaxy-galaxy lenses with MSE, so that the best possible subset can be studied in detail, is a key factor in pushing this field forward.

MSE will detect many lenses that are also identified in imaging surveys such as LSST. However, galaxy-scale strong gravitational lenses discovered by spectroscopic survey facilities such as MSE afford significant advantages over those identified in imaging surveys alone. First, the spectroscopic evidence of two objects along one line of sight eliminates the ambiguity of interpretation that can be associated with many imaging-selected lens candidates. Second, lenses that are selected through their spectra have known foreground and background redshifts immediately upon discovery, which is necessary to translate angular observables into mass measurements, and to quantify the integrated optical depth to low-mass CDM halos.

\subsection{Wobbling of the brightest cluster galaxies}\label{sec:bcg}

\begin{figure}
    \centering
    \includegraphics[width=0.48\textwidth,trim=0.1in 2.2in 0.05in 1.5in,clip] {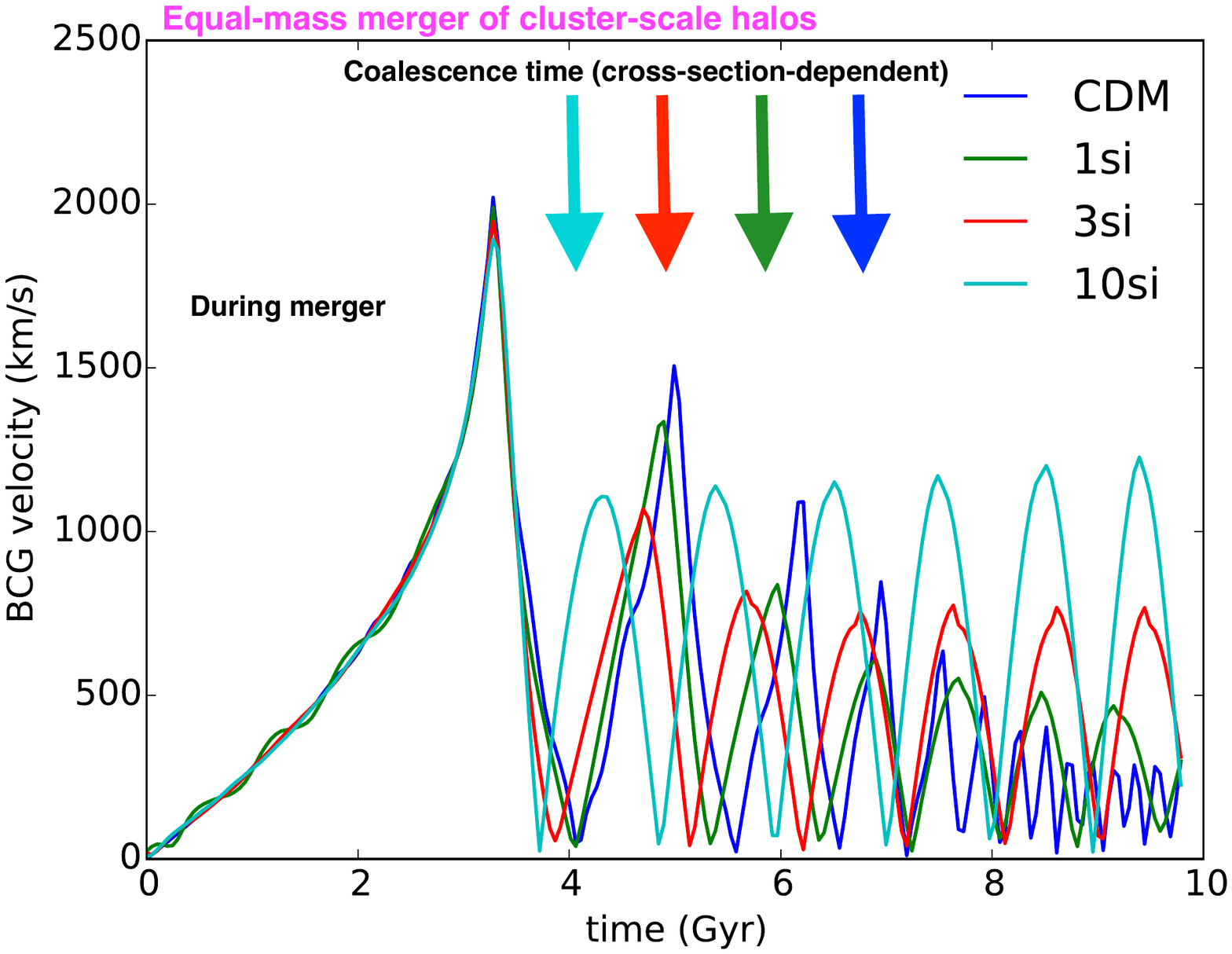}
    \includegraphics[width=0.50\textwidth]{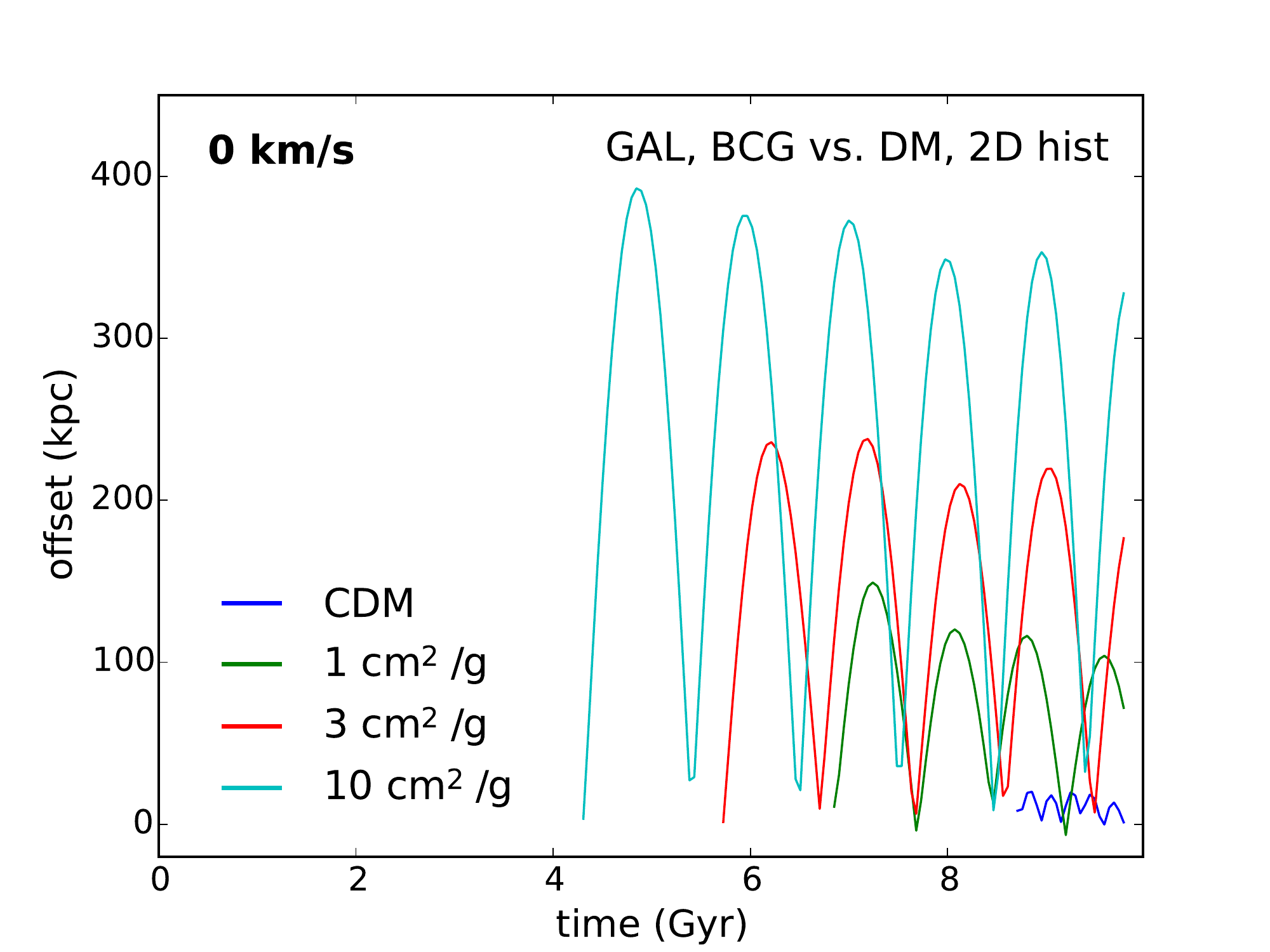}
    \caption{Left: Relative velocities of the two BCGs from an equal-mass merger of two $M_h = 10^{15}M_\odot$ cluster halos.  Simulations start at $t=0$ when the virial radii of the two halos touch, and free-fall on a radial trajectory toward each other. Right: Separation of the BCGs with the center of the merged cluster. Clusters are simulated with cross sections $\sigma/m = 0\hbox{ (CDM)}, 1, 3, \hbox{ and } 10 \hbox{ cm}^2/\hbox{g}$.  The time at which the clusters coalesce (the halo finder finds only one, not two, distinct halo centers) is marked with an arrow for each simulation.  The equal-mass merger simulations come from \citet{Kim:2016ujt}. Figure from Kim \& Peter, in preparation.}
    \label{fig:wobblingBCG}
\end{figure}

The hierarchical assembly of galaxy clusters leads to another probe of dark matter self-interactions, distinct from the density profiles illuminated by gravitational lensing.  Excitingly, strong SIDM cross section constraints may be obtained after the merger between two cluster-scale objects is complete, possibly stronger than those arising from phenomenology during the merger process.  In both cases, spectroscopy is essential for assessing the dynamical state of the system, a key ingredient in mapping cluster observables to SIDM cross section measurements.  Furthermore, in both cases, robust SIDM constraints will require a careful observational analysis of many systems matched with sophisticated, observationally-focuse, simulations. The era of ``easy'' SIDM constraints is over.

Consider the case where we catch a cluster during an accretion event.  Clusters consist of dark matter (halo of the host and subhalos containing galaxies), hot gas, and galaxies, in decreasing order of contribution to the total mass.  Each of these three components behaves differently during a merger.  Galaxies are effectively collisionless particles.  The hot gas is a highly collisional fluid.  In the CDM model, dark matter behaves like the galaxies, collisionlessly, although its dynamical evolution differs from that of galaxies on account of its different mass distribution throughout the object.  If dark matter is collisional, it will behave neither as galaxies nor as gas.  During the passage of a smaller halo (a cluster-, group- or galaxy-sized halo) through a cluster, the smaller halo will feel an extra force beyond gravitational if the self-scatter cross section is non-zero.  Non-gravitational interactions between host and subhalo particles can lead to a loss of specific momentum of the subhalo, either because of a ``drag"-like force \citep[typical for small-angle scattering, but it can also happen for large-angle scattering for close to equal-mass mergers;][]{markevitch2004,kahlhoefer2014,Kim:2016ujt,kummer2018} or because the tail of particles ejected from the subhalo can pull gravitationally on the subhalo \citep{Kim:2016ujt}.  The latter will affect the stellar component of the subhalo as well.  The former, though, is a force solely on the dark matter and not the baryonic components.  This may lead to a separation (``offset") between galaxies and dark-matter halos.

Spectroscopy is required to measure the kinematics of the merger as well as to obtain high-fidelity strong lensing maps. This is because offsets are expected to be smaller than typical strong-lensing uncertainty on the peaks of the dark matter density field and are velocity-dependent.  Historically, the spectacular Bullet Cluster merger has been the focus of SIDM constraints \citep[e.g.,][]{markevitch2004,clowe2006,randall2008,robertson2017,robertson2017b}, although more merging clusters with a variety of configurations are also being discovered and analyzed in the context of SIDM \citep[e.g.,][]{bradac2008,dawson2012,golovich2017a}.  Constraints based on simple analytic models of SIDM-induced offsets were typically of order $\sigma/m\lesssim 5 \hbox{ cm}^2/\hbox{g}$.  However, recent simulations show that offsets are much smaller than these simple analytic models suggest, are transient (largest offset just after pericenter passage and approaching the next pericenter), and depend on both the dynamics of the merger and the microphysical scattering model \citep{Kim:2016ujt,robertson2017,robertson2017b}.  Offsets are typically no greater than $\sim 20$ kpc for a hard-sphere cross section of $\sigma/m = 1\hbox{ cm}^2/\hbox{g}$, which is of order or smaller than typical uncertainties in the centroid of the subcluster galaxy distribution \citep[a problem intrinsic to the small number of confirmed galaxies even for relaxed clusters;][]{ng2017} and lensing peak positions.  More spectroscopy of more member galaxies and background lensed galaxies are critical to better constraining the merger dynamics, the centroid of the galaxy distribution, and the dark matter mass map from lensing.   

A recent analysis highlighted the importance of spectroscopy for measurements of ``bulleticity", the ensemble measurement of the offsets of many galaxy-mass subhalos infalling on cluster-scale halos \citep{massey2011,harvey2014,harvey2015}.  An initial detection of offsets in Abell 3827 (and a measurement of a cross section around $\sigma/m=1\hbox{ cm}^2\hbox{g}$ for large-angle scattering) was recently excluded on account of improved spectroscopy \citep[both optical and mm;][]{massey2015,kahlhoefer2015,massey2018}. 

Interestingly, competitive constraints on SIDM may arise from the positions and kinematics of galaxies in relaxed clusters.  In staged simulations of equal-mass mergers of cluster-sized halos, \citet{Kim:2016ujt} found that the orbits of galaxies at the centers of halos---notably the central galaxy, typically but not always the Brightest Cluster Galaxy (BCG)---continue to ``slosh" or ``wobble" about the relaxed cluster center after the merger (Figure \ref{fig:wobblingBCG}) for SIDM halos, but not CDM.  The origin of this effect is the inefficiency of dynamical friction for damping orbits in shallow gravitational potentials, and as such the amplitude and frequency of the oscillations about the relaxed cluster center (in position and velocity space) are expected to depend on the size of the SIDM core region. This in turn depends on the cross section.  \citet{harvey2017} claim a detection of BCG wobbling with respect to halo centers in position space.  However, the offsets have so far only been theoretically explored in detail in the context of equal-mass mergers \citep{Kim:2016ujt}, and are currently being investigated for more complex merger histories \citep{harvey2018}.  

An interesting approach is to look for relaxed remnants of equal-mass mergers (where we expect the wobbling to be greatest) by looking for systems with two bright central galaxies, and comparing the ensemble of position and velocity differences between the two central galaxies against simulations.  Observations of cluster gas and the kinematics of member galaxies can provide information on how relaxed the cluster is.  Exploring the separations in space and velocity of  systems with two bright central galaxies has the advantage of not requiring a $\sim 1-10~$ kpc-scale measurement of the halo center \citep{george2012,ng2017}.

\bibliography{references}

\end{document}